\ifx\LaTeXe\undefined
 old LaTeX version
\documentstyle[12pt,epsfig]{article}
  \newcommand{\ifig}[1]{\mbox{\epsfig{file=#1,width=10cm,angle=270}}}
\else
\documentclass[12pt]{article}
\usepackage{a4}
\usepackage[dvips]{graphicx}
\usepackage{latexsym}
\usepackage{euscript}
\usepackage{amsfonts}
\newcommand{\ifig}[1]{\includegraphics[height=8cm,width=14cm]{#1}}
\fi
\def\bc{\begin{center}}
\def\ec{\end{center}}
\def\be{\begin{equation}}
\def\ee{\end{equation}}
\def\bea{\begin{eqnarray}}
\def\eea{\end{eqnarray}}
\def\ba{\begin{eqnarray}}
\def\ea{\end{eqnarray}}
\newcommand{\amu}{A_{\mu}}
\newcommand{\m}{_{\mu}}
\newcommand{\VC}{\cal{V}}

\newcommand{\Coul} {\vec{\partial} \cdot \vec A }

\def\simge{\ \lower-
1.2pt\vbox{\hbox{\rlap{$>$}\lower5pt
\vbox{\hbox{$\sim$}}}}\ }

\newcommand{\AC} {{\cal{A}}}
\newcommand{\bibit}{\nineit}
\newcommand{\bibbf}{\ninebf}
\newcommand{\nonumsection}[1] {\vspace{12pt}\noindent{\tenbf #1}
        \par\vspace{5pt}}
\renewenvironment{thebibliography}[1]          
        {\ninerm 
         \baselineskip=11pt                             
         \begin{list}{\arabic{enumi}.}                  
        {\usecounter{enumi}\setlength{\parsep}{0pt}     
         \setlength{\leftmargin 17pt}{\rightmargin 0pt} 
         \setlength{\itemsep}{0pt} \settowidth          
        {\labelwidth}{#1.}\sloppy}}{\end{list}} 
\font\tenbf=cmbx10

 1
 1
 1
\font\ninebf=cmbx9
\font\ninerm=cmr9
\font\nineit=cmti9

\begin{document}
\pagestyle{empty} 
\vspace{3.6cm}
\vskip 2.0cm
\centerline{\large {\bf{Problems on Lattice Gauge Fixing }}}
\vskip 1.0cm
\centerline{L. Giusti$^{(1)}$, M. L. Paciello$^{(2)}$, C. Parrinello$^{(4)}$,  
S. Petrarca$^{(2,3)}$, B. Taglienti$^{(2)}$}
\vskip 10mm
\centerline{\small $^1$ Boston University - Department of Physics, 590 Commonwealth Avenue} 
\centerline{Boston MA 02215 USA.}
\centerline{\small $^2$ INFN, Sezione di Roma 1,
P.le A. Moro 2, I-00185 Roma, Italy.}
\centerline{\small  $^3$ Dipartimento di Fisica, Universit\`a di Roma ``La
Sapienza'',}
\centerline{\small P.le A. Moro 2, I-00185 Roma, Italy.}
\centerline{\small  $^4$ Synthesis Consulting}
\centerline{\small Centre International d'Affaires }
\centerline{\small 13 Chemin du Levant }
\centerline{\small 01210 Ferney-Voltaire, France.}
\centerline{\small }
\vskip 1.0in
\vfill
\newpage\clearpage
\pagenumbering{arabic}  
\setcounter{page}{1}
\tableofcontents
\newpage\clearpage

\pagestyle{plain}
\newpage 

  \setcounter{section}{0}
\setcounter{equation}{0}
\section{Introduction}
\label{sec:INTRO}
\indent Lattice gauge theories \cite{wilson74} are the only known rigorous 
non-perturbative formulations of non Abelian
theories and their interpretation as regularized versions 
of continuum quantum field theories in Euclidean space 
has been among the most fruitful theoretical ideas in 
the last years. They also offer the unique possibility to
compute correlation functions non-perturbatively 
through numerical simulations, and  therefore they represent  
a formidable laboratory where 
formal propositions may be tested and fundamental
phenomenological quantities can be computed from first principles.

Lattice gauge theories are defined on a discretized space-time which
cuts-off the high and low frequences and renders the theory finite. The
fundamental gauge fields are elements of the underlying group and,
since the group is compact, the lattice functional integrals
are well defined without any gauge-fixing.

In the following we mostly consider 
pure Yang-Mills theories (without fermions). Yet it is interesting to note 
that, if the fermionic operator satisfies the Ginsparg-Wilson relation
\cite{GWNEUB}, fermions can be introduced on the lattice preserving chiral 
and flavor symmetries at finite cutoff.

Non-perturbative lattice gauge-fixing becomes unavoidable to extract 
information from gauge-dependent correlators \cite{mandula99}. It is necessary
in order to study the propagators of the fundamental fields 
appearing in the QCD Lagrangian in the non-perturbative region.

It is also necessary in some non-perturbative renormalization 
schemes \cite{np,parrinello}
which use gauge dependent matrix elements to renormalize composite 
operators, and it can become a fundamental technical ingredient 
in the so called non gauge invariant quantizations of chiral
gauge theories \cite{romaappr}. These motivations justify the efforts
to obtain a consistent non-perturbative lattice gauge fixing.

In this review we will discuss some of the problems in lattice gauge
fixing, selecting the topics on the basis of their 
importance and of our personal
experience in the field.

In Sec. \ref{sec:LGF} we review the most popular approach to define
gauge-fixing on the lattice: we give the definitions
of the gauge dependent correlation functions and we sketch
briefly the steps of the numerical procedures adopted.

In Sec. \ref{sec:LACOUL}  we review the non-perturbative 
definition of the Landau and Coulomb gauges 
and we describe the main algorithms 
used in the literature to enforce these gauges numerically.

In Secs \ref{sec:SOFTGAUGE}, \ref{sec:GENERIC}, \ref{sec:LAPLACIAN},
\ref{sec:QUASITG}
 we review other lattice gauge conditions and
the corresponding gauge-fixing procedures proposed in the literature. 

In Sec. \ref{sec:CONFINEMENT} we briefly sketch the r\^ole of the gauge choice
in understanding the physics of quark confinement. 
 
Sec. \ref{sec:LANGE} is devoted to the gauge fixing implementation 
in the Langevin dynamics algorithm, which is necessary to overcome 
divergent fluctuations along the gauge directions.

In Sec. \ref{sec:LGP} we discuss some problems related to the ambiguities in the lattice 
definition of the gauge potential. Different regularized definitions 
and their effects on gauge dependent quantities are analyzed. 

In Sec. \ref{sec:GRIBOV}   we discuss the problem of 
numerical Gribov copies, their effects on physical quantities
and some approaches which have been proposed to remove this ambiguity.

Sec. {\ref{sec:SMOO}}~~is devoted to the lattice QCD 
gauge dependent smoothing procedures.  

In Sec. \ref{sec:CONCLUSIONS} we draw our conclusions and acknowledgements.

  \section{Standard Non-Perturbative Gauge Fixing}\label{sec:LGF} 
In the standard formulation of lattice gauge theories
proposed by Wilson~\cite{wilson74} the
link $U_\mu(x)$ are the fundamental gauge fields of the theory, they
are group elements of $SU(N)$ in the fundamental (N-dimensional)
representation and they transform under a gauge transformation $G(x)$
as 
\begin{equation} \label{eq:tras}
 {U^{G}_\mu (x)}=G(x) U_\mu (x) G^{\dagger} (x+\mu).
\ee
The gauge invariant action
is defined as     
\be   
S = \beta \sum_{plaq} \Big[ 1 - \frac{1}{2N}   
\mbox{Tr} \left[ P_{\mu\nu}(x) + P^{\dagger}_{\mu\nu}(x) \right] \Big]   
\label{eq:sg}   
\ee   
where, in the standard notation, $\beta=2N/g_0^2$, $g_0$ is the bare 
coupling constant, $P_{\mu\nu}(x)$ is the Wilson plaquette, i.e. 
the path-ordered product of link variables  
\be
P_{\mu\nu}(x) = U_\mu(x) U_\nu(x+\mu)U^{\dagger}_\mu(x+\nu) U_\nu^\dagger(x) 
\ee 
around the boundary of a plaquette $P$. 
The expectation value
of any gauge invariant operator ${\cal O}(U)$ is given by
\ba\label{eq:averages}
\langle {\cal O} \rangle & = & \frac{1}{Z}\int dU e^{-S(U)} {\cal O}(U)\\
Z & = &  \int dU e^{-S(U)}\nonumber\; ,
\ea 
where $dU$ denotes the group-invariant integration measure over the links
satisfying the following properties:
\ba\label{eq:GImeasure}
\int h(U) dU = \int h(VU) dU & = & \int h(VU) dU \qquad \forall \;V , U\in G\;\\
\int dU & = & 1\nonumber\; ;
\ea
being $h(U)$ a generic function of the links.
Since the domain of the link integration is compact, the lattice functional 
integrals~(\ref{eq:averages}) are well defined and the gauge invariant 
correlation functions can be computed without fixing the gauge.

In the ideal case where the gauge-fixing 
condition $f(U^G)=0$ has an unique solution for 
each gauge orbit, i.e. there are no Gribov copies 
\cite{Gribov}, the Faddeev-Popov
procedure \cite{FP} can be applied. The 
gauge-invariant Faddeev-Popov 
determinant $\Delta_f(U)$
is defined as 
\be\label{eq:FPdeterminant}
\Delta_f(U) \int dG \delta(f(U^G)) = 1\; ,
\ee 
where the integration is over all gauge transformation $G$. 
By inserting  the previous identity in the functional integrals
(\ref{eq:averages}), changing the variables $U \rightarrow U^{G^{-1}}$
and the order of integration, using the 
gauge invariance of $dU$, $S(U)$, 
$\Delta_f$ and of ${\cal O}(U)$, we can write
\be\label{eq:omedio}
\langle {\cal O} \rangle_f =  \frac{1}{Z}\int dU e^{-S(U)}
\Delta_f(U)\delta(f(U)) {\cal O}(U)\;.
\ee
For gauge-invariant quantities this expression is equivalent to
Eq.~(\ref{eq:averages}). Eq.~(\ref{eq:omedio})
is the Faddeev-Popov definition of the correlation functions
of gauge-dependent operators. 

In absence of Gribov copies, the Faddeev-Popov
determinant can be expressed as an integral 
over the ghosts and anti-ghost fields $\eta$ and ${\bar{\eta}}$, obtaining
\be\label{eq:magari}
\langle {\cal O} \rangle_f = \frac{\int dU d\lambda d\bar \eta d\eta\;\; 
 e^{-S(U)} e^{-\frac{\alpha}{2}\int \lambda^2}e^{\delta \int f \bar \eta}{\cal O}(U)}
{\int dU d\lambda d\bar \eta d\eta \;\; e^{-S(U)} e^{-\frac{\alpha}{2}
\int \lambda^2}e^{\delta \int f \bar \eta}}
\ee
where $\lambda$ 
are Lagrangian multipliers  and $\delta$ represents the lattice BRST~\cite{BRST}
transformations defined as 
\ba\label{eq:BRSTlattice}
\delta U_\mu  & = & \eta(x) U_\mu(x) - U_\mu(x) \eta(x+\mu)\nonumber\\
\delta \bar \eta & = & i\lambda\\
\delta \eta & = & \frac{1}{2}\eta \eta \nonumber\\
\delta \lambda &= & 0\; .
\ea
The gauge-fixed action in (\ref{eq:magari}), including the ghost terms,
is local and is invariant under BRST transformations.
The Faddeev-Popov procedure can be replaced by the more formal apparatus of the 
BRST symmetry. 
It resembles the 
continuum formula
\begin{equation} \label{eq:omediocont}
\langle {\cal O} \rangle = 
\int \delta\amu \delta \eta \delta \bar{\eta}\, {\cal O}\, e^{-S(A)-
S_{ghost}(\eta,\bar{\eta},A)} \delta(f(A))\; .
\; ,\nonumber
\ee

On the contrary in presence of Gribov copies, i.e. multiple 
solutions of the equation
\be\label{eq:maledetta}
f(U^G)=0
\ee
for a given gauge configuration $U$,
the Faddeev-Popov determinant cannot be expressed as an integral over 
the ghosts and the BRST invariance is lost. 
Labelling the different solutions of~(\ref{eq:maledetta})
by $G_i$, in the integral functional (\ref{eq:omedio}) we sum over 
several gauge equivalent copies of the 
same configuration obtaining
\be\label{eq:maledetta2}
\Delta(U)^{-1} = \sum_i \frac{1}{|\mbox{det}\frac{\delta f(U^G))}{\delta
    G}|_{G=G_i}}\; .
\ee
Eq.~(\ref{eq:maledetta2}) leads to an acceptable but very
inconvenient gauge fixing procedure. 

Alternative procedure to mantain the BRST symmetry in the gauge 
fixing process despite of the presence of the Gribov copies have been
proposed~\cite{sharpe}. But Neuberger has shown~\cite{neuberger} 
that on the lattice the requirement of the standard BRST invariance 
of the gauge-fixed action leads to the non-perturbative level 
to disastrous results, i.e. the physical observables are reduced to the 
an undetermined form.
The argument can be sketched as follows~\cite{neuberger}: 
let us define the function
\be
F_{\cal O}(t) = \int dU d\lambda d\bar \eta d\eta \; e^{-S(U)} 
e^{-\frac{\alpha}{2}\int \lambda^2}e^{t \delta \int f \bar \eta} {\cal O}(U),
\ee
which satisfy
\ba
\frac{d F_{\cal O}(t)}{dt} = \int dU d\lambda d\bar \eta d\eta\;
\delta\left[ \int f \bar \eta \;e^{-S(U)} e^{-\frac{\alpha}{2}\int
  \lambda^2}e^{t \delta \int f \bar \eta} {\cal O}(U)\right] = 0
\ea 
because the integral of a total BRST variation vanishes identically.
On the other hand  $F_{\cal O}(0)=0$
because the integrand does not contain ghosts and therefore 
$F_{\cal O}(1)=0$ and $\langle {\cal O}\rangle = \frac{0}{0}$.
Possible ways out of this paradox have been proposed in \cite{massimo,massimo1} 
by imposing a modified BRST symmetry on the 
lattice which converges towards the conventional one in 
the continuum limit, interesting  
remarks on this subject can also be found in~\cite{fuji}.

The standard numerical gauge-fixing procedure on the lattice~\cite{mandu87} is
obtained by reversing the Faddeev-Popov analysis described above and
neglecting the presence of Gribov copies. 
In ~(\ref{eq:omedio}), by multipling by $\int dG$ (nothing depends on $G$),
changing the order of integration, changing variables $U \rightarrow U^{G}$, 
using the gauge invariance (\ref{eq:GImeasure}) of $dU$, $S(U)$,
and performing the integral over $G$ (taking  
into account the $\delta$ function) we can write
\be\label{eq:algorithm}
\langle {\cal O} \rangle_f =  
\frac{1}{Z}\int dU e^{-S(U)} {\cal O}(U^{G(U)})\; ,
\ee
where $G(U)$ is the gauge transformation for which $f(U^{G(U)})=0$.
The definitions in Eqs.~(\ref{eq:omedio}) and (\ref{eq:algorithm})
of the correlation functions of gauge dependent operators are 
equivalent if the equation $f(U^G)=0$ has a unique solution,
i.e. there are no Gribov copies.
In this case the Faddeev-Popov determinant $\Delta_f(U)$ cancels out
when the integral of the $\delta$ function is performed. 

The Eq.~(\ref{eq:algorithm}) summarizes the procedure
implemented by the numerical algorithm~\cite{mandu87}: 
\begin{itemize}
\item  a set of $N$ thermalized configurations 
${\{\cal C}_U\}$ is generated with periodic boundary conditions 
according to the gauge invariant weight $e^{-S(U)}$;
\item  for each  $\{{\cal C}_U\}$ a numerical algorithm computes 
the gauge transformation $\{G\}$;
\item each thermalized configuration $\{{\cal C}_U\}$ is gauge rotated
 obtaining the set of the gauge fixed thermalized configurations of the
 gauge-fixed set $\{ {\cal C}_{U^G}\}$;
\item the expectation value of an operator is
 given by the average of the values of the operator evaluated
 at each  gauge rotated configuration of the gauge-fixed set: 
\end{itemize}
\be
\langle{\cal O}\rangle^{Latt}=\frac{1}{N}\sum_{\{conf\}}{\cal O}({U}^G)\;\; .
\ee
The complexity of the ghost technique is replaced
by the large amount of computer time spent 
to obtain numerically the gauge transformations which satisfy the
gauge condition required. The entire procedure is rigorous only when the
gauge fixing condition is free from Gribov copies. Otherwise 
the effect of Gribov copies
must be taken into account (see Section~\ref{sec:GRIBOV}),
 since the definition of the correlation functions would depend 
on the way the gauge fixing algorithm selects a preferred Gribov copy.

In the case of an imperfect or inadequate  gauge fixing the 
measurement of a gauge dependent operator is at best 
affected by additional fluctuations to be summed up  to the
intrinsic statistical noise~\cite{papape92}. In other cases,
as in some calculations on $U(1)$~\cite{plewnia,defo91,mitri93}
 and on confinement 
vortex picture~\cite{borko21,koto2}, the influence of lattice Gribov 
copies can mask the regular behaviour of a measurement (see later).

  \section{Landau and Coulomb Lattice Gauge-Fixing}
\label{sec:LACOUL}
The standard way of fixing the Coulomb and Landau gauges on the 
lattice~\cite{wil79, Wil80, mandu87, davi, berna90} is based 
on the numerical minimization of the functional 
\begin{equation}  
F_U[G]= -Re \; Tr \;\sum_x \sum_{\mu=1}^l {U_{\mu}}^{G(x)} (x)\; ,
\label{eq:effel}
\end{equation}
where $l$ is $3$ for Coulomb and $4$ for Landau gauge.
$F_U[G]$ is constructed in 
such a way that its extrema $G^*$
\be
\left. \frac {\delta F} {\delta G}\right\vert_{G^*} =0
\ee
are the gauge fixing transformations 
corresponding to the discretized gauge condition 
\begin{equation} \label{eq:landlat}
\Delta^G(x) \equiv \sum_{\mu} \ ( A^G_{\mu} (x) - 
A^G_{\mu} (x - \hat{\mu} ))= 0 \; , 
\end{equation}
where
\be \label{eq:standard}
A_{\mu} (x) 
\equiv  \left[{{U_{\mu} (x) - U_{\mu}^{\dagger} (x)}\over
{2 i a g_0}}\right]_{Traceless}\; .
\ee 
The ``standard definition'' (\ref{eq:standard}) of the gauge 
potential is na{\"\i}vely suggested by the interpretation 
of $U_{\mu} (x)$ as the lattice parallel transport operator and 
by its formal expression in terms of the ``continuum'' gauge 
field variables, ${\cal A}_{\mu} (x)$ as:
\be
U_{\mu}(x)=e^{iag{_0} {\cal A}_{\mu} (x)}\; 
\label{eq:seconda}
\ee
where $a$ is the lattice spacing. The second variation of $F_{coul}$ is  
the lattice Faddeev-Popov operator for the Landau and Coulomb gauge.

The lattice gauge-fixing sketched above is the 
analogous of the continuum Gribov's procedure.
Moreover there is also a corrispondence between the lattice 
functional in Eq.(\ref{eq:effel}) and the continuum one
 \begin{equation} 
F_A [G] \equiv -\ Tr \ \int d^4 x \ \left( A_{\mu}^G (x) A_{\mu}^G (x) \right) 
\equiv -\left( A^G, A^G \right) \equiv -||A^G||^2 \; ,
\label{eq:effe}
\end{equation}
which reaches its extrema when the gauge-fixing condition
\be
\partial_\mu A^G_\mu=0, \; \;{\rm supplemented}\;{\rm with} 
\;{\rm periodic}\; {\rm boundary}\; {\rm condition} ;
\label{eq:landau}
\ee
is satisfied. $\mu$ goes from 1 to 3 or 4 in the case
of Coulomb and Landau gauge respectively.

It is remarkable that Eq.~(\ref{eq:effel}) does not 
correspond to the natural discretization of the continuum 
functional (\ref{eq:effe})
according to the lattice definition of the gluon field 
(\ref{eq:standard}) but it differs from that by $O(a)$ terms.
The form in Eq.~(\ref{eq:effel}) is adopted not only for 
its simplicity but also because it leads to the gauge condition
(\ref{eq:landlat}).

The most na{\"\i}ve algorithm to minimize the functional
in Eq.~(\ref{eq:effel}) sweeps the lattice
by imposing the minimization requirement one site 
at a time, and repeating the process until the gauge 
transform has relaxated  sufficiently 
into a minimum~\cite{wil79, mandu87}.
Actually in order to fix the gauge one needs just to reach any 
stationary point of $F$, hence the
requirement of a minimum is a  somewhat stronger 
request naturally adopted by the
numerical procedure.  

In order to study the approach of the functional (\ref{eq:effel}) 
to a minimum, the values of two quantities are usually numerically monitored. 
The first one is $F_U[G]$ itself, which decreases monotonically 
and eventually reaches a plateau. The other one, denoted by $\theta$, 
is defined as follows:
\begin{equation} 
\theta^G \equiv {1 \over V} \ 
\sum_{x} \theta^G(x) \equiv {1 \over V}
\ \sum_{x} Tr \ [ \Delta^G (x) (\Delta^G)^{\dagger} (x)],
 \simeq  \int d^4 x \  Tr(\partial_\mu A_{\mu}^G )^2 ,
\label{eq:thetalat}
\end{equation}
where $V$ is the lattice volume.
The function $\theta$ is a measure of the first derivative of 
$F_U[G]$ during the gauge-fixing process; it decreases (not 
strictly monotonically) approaching zero when $F[U^G]$ reaches 
its minimum. The desired gauge fixing quality is 
determined by stopping the computer code when $\theta^G$ has 
achieved a preassigned value close to zero which is often 
defined  the gauge fixing quality factor. 
In Fig.~\ref{thetaef} it is shown the typical behaviour of 
$\theta$ and of the quantity $ {F_r=|F -F_{min}|}$
as function of the gauge fixing sweeps. These behaviours can change quite
a lot among different thermalized configurations.
\begin{figure}      
\hspace{2.0cm}  
\includegraphics[height=150mm,width=120mm]{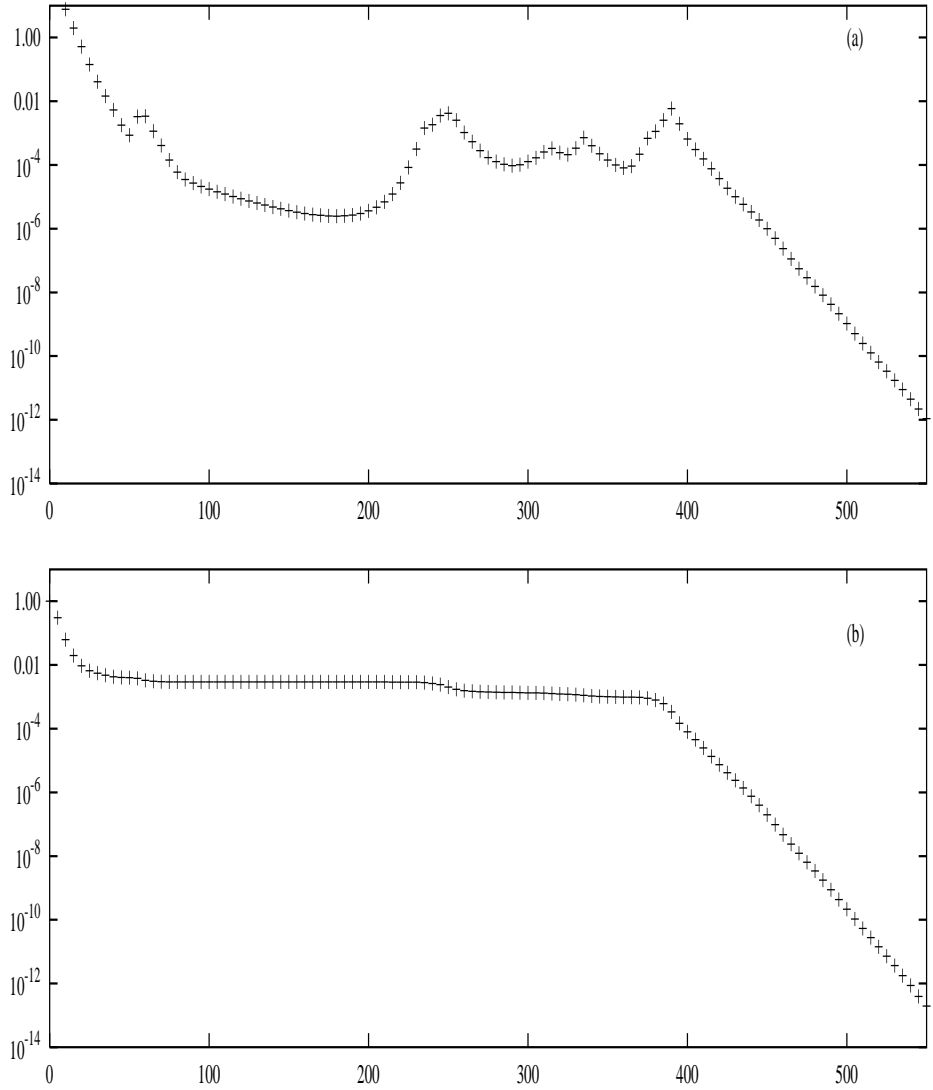}  
 \caption{Typical behaviour of $\theta$ (curve a) and
  $F_r$ (curve b) as function of the number of gauge fixing sweeps
   for the Landau gauge. Lattice size is $8^3 \cdot 16$.} 
 \label{thetaef}
\end{figure}

The choice of the gauge fixing quality is a delicate point in 
the case of a simulation with a large volume and a high number 
of thermalized configurations. 
Of course, the better is the gauge fixing quality, the more
computer time is needed. Moreover it is impossible to know 
in advance, before computing the gauge dependent correlation functions, 
whether an a priori criterion is suitable or not.
So that, the stopping $\theta$ value is normally fixed 
on the basis of a compromise between the available
computer time and the gauge fixing quality. Sometimes, in 
the case of calculations performed on computers with
single precision floating point, the maximum gauge fixing 
quality is limited by a value of the order of the floating 
point zero: $\theta \simeq  10^{-7}$, this value is usually 
enough to guarantee the stability of  gauge dependent correlators
even in high precision measurements, like the calculations of 
gluon and quark propagators (see for example Ref. \cite{mandula99})
and the calculus of the running QCD coupling \cite{heparpit95}
by means of the 3-gluon vertex function. In these cases gauge fixing 
becomes a time consuming part of 
the computation, comparable to the calculation of a quark propagator.

The lattice Landau and Coulomb gauge-fixing are 
affected by the problem of lattice Gribov copies. 
Section~\ref{sec:GRIBOV} is devoted to this issue. 

It is also interesting to see the gauge-fixing procedure from 
another point of view: it can be considered as
the process of finding the ground state of a dynamical
system (a spin system like in the Ising model) 
where $F$ takes the place of the hamiltonian, the gauge transformations
$G$'s are dynamical variables, belonging to the $SU(3)$ group,
and the links are the couplings. This analogy is clearly seen
by writing the gauge trasformation~(\ref{eq:tras})    in  
the functional form~(\ref{eq:effel}).

\subsection{Acceleration of Landau and Coulomb Gauge-Fixing}
For large lattices the gauge fixing algorithms, as other iterative methods, 
converge slowly due to long-range 
correlations and large condition numbers of the matrices 
which control the algorithms. This is a crucial problem usually called 
{\it critical slowing-down} (see for example \cite{wolfsok9091}).
To reduce the critical slowing-down in 
gauge fixing algorithms, two classes of improvements are often adopted: 
\begin{itemize}
\item overrelaxation, originally proposed in Ref.~\cite{mandu90},   
to speed up gauge-fixing algorithms;
\item Fourier preconditioning~\cite{davi} to adjust the 
matrix governing the system evolution so that all the
eigenvalues become approximately equal to the largest 
one without affecting the final answer.
\end{itemize}
The overrelaxation algorithm, a technique originally introduced 
to improve the convergence of iterative methods to solve 
classical linear algebra and differential equation problems, 
is particularly suitable to face critical slowing down in 
numerical simulations, as first shown in Ref.~\cite{adler81}. 
The effectiveness of this method has been studied in different 
papers; for a general discussion see also 
Refs.~\cite{adler81,adler84,good86,creutz87,
brown87,adler88,deforcra89,papar92}.
The  overrelaxation method is implemented 
in the process of gauge fixing by 
replacing $G(x)$ with its power $G^{\omega}(x)$, 
at each iteration. In practical computations, $G^{\omega}(x)$ is given by a
truncated binomial expansion
\begin{equation}  
G^{\omega} = \sum_{n=0}^N \frac{\gamma_n(\omega)}{n!}{(G-I)}^n \, ,
\quad
\gamma_n(\omega) = \frac{\Gamma(\omega + 1)}{\Gamma(\omega + 1 - n)}
\label{eq:o3}
\end{equation}
where $1 <\omega <2$ and usually $2<N<4$.
The $\omega$ parameter is tuned empirically at an optimal 
value $\omega_{opt}$, (a typical value is $\omega_{opt}\simeq 1.75$),
to reach the fastest convergence. Before $G^{\omega}(x)$ is applied
on the link, it has to be appropriately normalized to belong to 
the gauge group.  

The study of the convergence as a function of the overrelaxation 
parameter is shown in Fig.~\ref{F_Mandu}.
\begin{figure}      
\hspace{2.0cm}  
\includegraphics[height=80mm,width=120mm,]{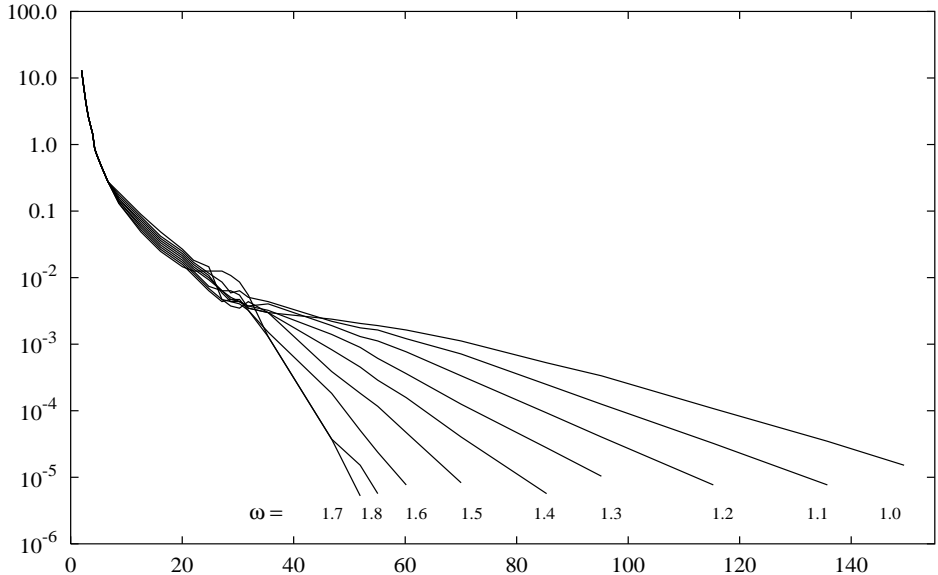}   
 \caption{Efficacy of gauge fixing as a function on the 
overraxation parameter $\omega$ from
Ref.~\cite{mandu90}. The plot shows the decrease of $\theta$ for different
values of $\omega$ and a $4^4$ lattice at $\beta =5.6$.} 
 \label{F_Mandu}
\end{figure}
It shows clearly that 
it is possible to distinguish two different 
regimes~\cite{mandu90,adler81,adler84,papar92}.
The initial behaviour appears to be almost insensitive to the value of 
~$\omega$ and corresponds to large local fluctuations 
 from one iteration to another. In the second 
regime the fluctuations 
are smooth, characterized by a long-range pattern, and the rate 
of convergence is quite sensitive to the $\omega$ value since 
here one faces the problem of critical slowing down and the 
optimization of the algorithm becomes crucial.
On the lattices considered in Ref.~\cite{papar92}  a 
factor of  $3$ or $5$ is gained in the number of iterations
while a computer time overhead of a factor $1.5$ 
for iteration is paid so that the resulting gain in time due to this 
procedure\cite{papar92} is about a factor of $2$.
Another small price to pay is related to the need for tuning 
empirically $\omega$ at an optimal value $\omega_{opt}$ for which 
convergence is most rapid, fortunately this value does not
depend too much~\cite{papar92}
on different configurations at fixed $\beta$ and $V$.
It is remarkable that any algorithm for Landau or Coulomb 
gauge fixing can be easily modified in such a way to include
the overrelaxation, the only change required is 
to add the expansion (\ref{eq:o3}) at the end of each iteration 
of the normal gauge fixing procedure algorithm.
A variation of the overrelaxation method is the stochastic 
overrelaxation algorithm, proposed by Ref.~\cite{deforcra89}, 
in which a local gauge transformation $G(x)^2$ replace $G(x)$ 
with probability $p$. The actual acceleration gain turns out 
to depend strongly on $p$ and the procedure definitely diverges 
for $p = 1$.

Monitoring the deviation from the Landau gauge at each site,
it is possible to see a broad range of small deviations while 
in the Fourier space many  slowly decaying modes are observed, including,
but not limited to, the longest wavelenghts.     

The Fourier acceleration (FA) technique applied to the gauge fixing, 
alleviates the problem of critical 
slowing down. The idea is to precondition 
the problem using a diagonal matrix in momentum space which is 
related to the solution of a simplified version of the 
problem \cite{davi,batro85}.
The overall relaxation time is determined 
by the smallest eigenvalue in the momentum space of the matrix 
governing the iterative algorithm. 

Here  we  describe the use  of Fourier acceleration to improve 
the convergence of the  the Landau gauge fixing algorithm, 
following Ref.~\cite{davi}. This method modifies the 
gauge transformation $G(x)$ in Fourier space  in such a way that 
all modes converge as fast as the fastest mode.
In order to fix a lattice version of Landau gauge one minimizes the 
expression (\ref{eq:effel}) in the space of gauge-equivalent field
${U_{\mu}}^{G}$ defined in relation~(\ref{eq:tras}).
Following a na{\"\i}ve steepest-descent method, we differentiate
with respect to the gauge trasformation, and at each step 
of the iterative procedure, $G(x)$ takes the expression:
\be\label{eq:four}
G(x)= exp\bigg\{\frac{\alpha}{2}\bigg[\sum_{\nu}\bigg[\Delta_{-\nu}(U_{\nu}(x)
-U_{\nu}^{\dagger}(x))-1/N_C\mbox{\rm Tr}[\Delta_{-\nu}(U_{\nu}(x)
-U_{\nu}^{\dagger}(x))]\bigg]\bigg]\bigg\}
\ee
where $\Delta_{\nu}(U_{\mu}(x)) = U_{\mu}(x - \hat{\nu}) - U_{\mu}(x)$
and $\alpha$ is a tuning parameter.
To optimize the convergence, the Fourier accelerated method 
replaces the Eq.~(\ref{eq:four}) by
\be
G(x)= exp\bigg[\widehat{F}^{-1}\Big( \frac{\alpha}{2}\frac{p^2_{max}a^2}
{p^2a^2}\widehat{F}\Big(\sum_{\nu}\Delta_{-\nu}
(U_{\nu}(x)-U^{\dagger}(x))-trace\Big) \Big)\bigg]
\ee
where $p^2$ are the eigenvalues of the lattice version of the 
$(\partial^2)$ operator, $a$ is the lattice spacing and 
$\widehat{F}$ denotes the Fourier transform operator. Thus, 
in this case, the preconditioning is obtained using in momemtum 
space a diagonal matrix with elements given by $1/p^2$ \cite{davi}.
The FA method in its original form makes use of the fast 
Fourier transform algorithm to evaluate $\widehat{F}$ and $\widehat{F}^{-1}$, 
which requires a computer time proportional to $VlogV$ where $V$
is the lattice 
volume \cite{davi}, making it very appealing from the numerical 
point of view. Note, however, that on parallel machines the cost of the 
Fourier transform is not negligible because of its high non-locality.
For this reason the FA is not considered appealing anymore in large 
scale simulations. An implementation of the FA method for Landau gauge 
fixing, avoiding completely the use of the Fourier transform, has been 
proposed and tested for the 4-dimensional $SU(2)$ case, on serial 
and on parallel machines, in Refs. \cite{cucchie98,cucchie97}.
In Figs.~\ref{F_Four1} and \ref{F_Four2} the behaviours 
of a gauge fixing evolution with and without the Fourier accelerated 
algorithm as a function of the iteration number are shown.
\begin{figure}      
\hspace{2.0cm}  
\ifig{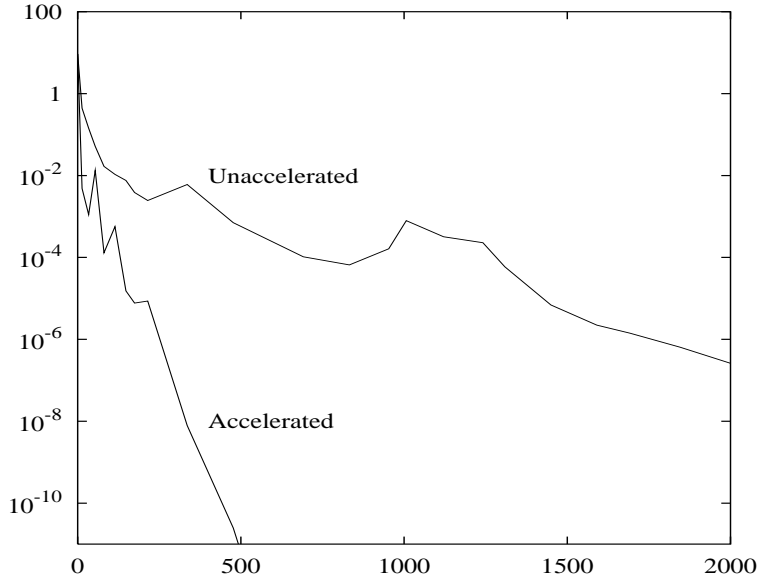}  
 \caption{ $\theta$ plotted as a function of iteration number 
for gauge fixing with and without Fourier acceleration, redrawn from
  Ref.~\protect\cite{davi}.} 
 \label{F_Four1}
\end{figure}
\begin{figure}      
\hspace{2.0cm}  
\ifig{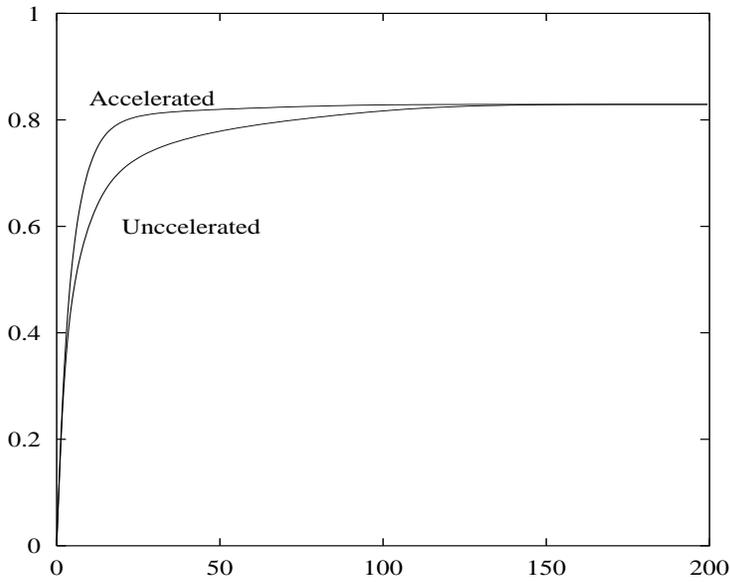}  
  \caption{ $F_U$ (only the first 200 steps
  are shown) plotted as a function of the gauge fixing iteration number 
for the same configuration as in Fig.~\ref{F_Four1} with 
and without Fourier acceleration, redrawn from
  Ref.~\protect\cite{davi}.} 
  \label{F_Four2}
\end{figure}

In Ref.~\cite{mandu99} global gauge fixing on the lattice specifically 
to the Landau gauge, is discussed with the goal of understanding 
the question of why the process becomes extremely slow for large lattices.
The author constructs an artificial ``gauge fixing'' problem which has the 
essential features encountered in the real case. In the limit in which the 
size of the system to be gauge fixed becomes infinite, the problem becomes 
equivalent to finding a series expansion in functions which are related 
to the Jacobi polynomials. The series converges slowly, as expected. 
It also converges non-uniformly, which is an observed characteristic of 
gauge fixing. In the limiting example of Ref.~\cite{mandu99} the 
non-uniformity arises through the Gibbs phenomenon.

Gauge fixing algorithms have also been used as a prerequisite 
for the Fourier acceleration of other procedures required for 
simulations of lattice gauge theories as the matrix inversion, 
for algorithms based on the Langevin equation and others.

A review of the performances of some standard gauge procedures 
for non-Abelian gauge theories can be found in Ref. \cite{sushi93}. 
Techniques involved in gauge fixing, as reunitarization and convergence 
criteria, accelerating procedures and performance of algorithms on 
the parallel machines $CM2$ and $CM5$ are also analyzed.
Critical slowing-down of several gauge fixing algorithhms for 
different gauges in the $SU(2)$ case, at zero temperature, is
discussed in Ref. \cite{cucchie99}.

  \section{The Soft Covariant Gauge}
\label{sec:SOFTGAUGE}

A nonperturbative method for gauge fixing in the continuum has been proposed
by Jona-Lasinio and Parrinello \cite{jlp} and by Zwanziger 
\cite{zwanziger} and extended on the lattice in \cite{ioeste}.
They suggest to modify 
the gauge invariant Wilson's partition function
\begin{equation}
Z =  \int d U \ e^{- S(U)},
\label{eq:Wilson}
\end{equation}
 by simply inserting an identity in it
(\ref{eq:Wilson}):
\begin{equation}
 Z_{mod} \equiv
\int d U \ e^{- S(U)} \ I^{-1} [U] \ \int dG \ e^{- \beta M^2
F [U^{G}]}. 
\label{eq:zmod}
\end{equation} 
$F [U^{G}]$ is a generic function of the links 
not invariant under general gauge transformations and  
the gauge invariant quantity $I [U]$ is given by:
\begin{equation}
I [U] \equiv
\int dG \  e^{- \beta M^2 F [U^{G}]}.
\label{eq:ilat}
\end{equation}
Eq.~(\ref{eq:zmod}) corresponds to
 the first step of the standard Faddeev-Popov gauge-fixing procedure. 
However, unlike what would happen in the continuum, 
 $Z_{mod} \equiv Z$ is a finite quantity because the group of gauge 
transformations on a finite
lattice is compact. Thus   
$Z_{mod} $ can provide a new definition for 
 the expectation value of
 $O [U]$:
\begin{equation}
< O >_{mod} \equiv Z_{mod}^{-1} \ \int d U \ e^{- S(U)} \ I^{-1} [U] \ \int dG \ 
e^{- \beta M^2 F [U^{G}]} \ O [U^{G}].
\label{eq:mediamod}
\end{equation}    
\noindent
If $O [U]$ is a gauge invariant operator then $ < O > =
< O >_{mod} $, while eq.~(\ref{eq:mediamod})
defines the expectation value of gauge-dependent operators.

By defining
\begin{equation}
< O [U] >_{G} \equiv I^{-1} [U] \ \int dG \ 
e^{- \beta M^2 F [U^{G}]} 
\ O [U^{G}], 
\label{eq:mediaG}
\end{equation}
$< O >_{mod}$ can be cast in the form:
\begin{equation}
< O >_{mod}
 = {\int d U \ e^{- S(U)} \ < O [U] >_{G} \over 
\int d U \ e^{- S(U)}} = < \ 
< O [U] >_{G} 
\ >.
\label{eq:final}
\end{equation}
The above expression indicates that in the gauge-fixed model the 
expectation value of a gauge dependent quantity $O [U]$ 
is obtained in two steps. 
First one associates with $ O [U] $ the gauge 
invariant function 
$ < O [U] >_{G} $, which has the form of a Gibbs average of $ O [U^{G}]$ over 
the group of gauge transformations, with a statistical weight factor 
$e^{- \beta M^2 F [U^{G}]}$.
Then one takes the average of $ < O [U] >_{G} $ {\it \`a la} Wilson. 
\par 
This suggests the following numerical algorithm: 
\begin{itemize}
\item generate a set of link configurations $ U_{1}, \ldots 
U_{N} $, weighted by the Wilson action, via the usual gauge 
invariant Monte Carlo algorithm for some value of $ \beta $; 
\item use each of the $U_{i}$
 as a set of quenched bonds in a new 
Monte Carlo process, where the dynamical variables are the 
local gauge group elements $ G(x) $ located on the lattice sites.
These are coupled through the links $ U_{i}$, 
according to the effective Hamiltonian $F[U_{i}^G]$.
In this way one can produce for every link configuration $ U_i$
an ensemble of gauge-related configurations, weighted by the 
Boltzmann factor $exp (- \beta M^{2} F [U_{i}^{G}])$.
We call $ < O [U_i] >_{G} $ the average of a gauge dependent 
observable $O$ with respect to such an ensemble, in the spirit of
Eq.~(\ref{eq:final}).
\item 
Finally, the expectation value 
$< O >_{mod}$
is simply obtained from the Wilson average of the $ < O [U_i] >_{G} $, i.e.: 
\begin{equation}
< O >_{mod} \approx {1 \over N} \ \sum_{i = 1}^{N} \  < O [U_i] >_{G}. 
\protect\label{eq:finmod}
\end{equation}
\end{itemize}
In the above scheme $M^2$ can be interpreted as 
a gauge parameter, which determines 
the effective temperature $1 / \beta M^{2}$ 
of the Monte Carlo simulation over the group of gauge transformations.

Adopting as gauge fixing action the form of the Landau 
gauge fixing functional~(\ref{eq:effel}) it is possible to 
get the connection between this scheme and the
usual Landau gauge fixing.
It turns out that the stationary points of $F [U^G] $ correspond to
link configurations $U^G$ that satisfy the lattice version of the Landau gauge
condition. All such configurations correspond to Gribov copies. In particular, 
those corresponding to local minima of $F [U^G] $ also satisfy a positivity
condition for the lattice Faddeev-Popov operator \cite{Zwa91}. As a
consequence, in the limit $ M^{2} \rightarrow \infty $,  the above gauge-fixing
is equivalent to the so-called minimal Landau gauge condition, which prescribes
to pick up on every gauge orbit the field configuration corresponding to the
absolute minimum of $F [U^G] $ \cite{ZWA2}.  

While this method is conceptually very simple, 
much less is known about its perturbation theory expansion, 
as compared with the standard gauge fixing based on BRST invariance.
In fact in the Faddeev-Popov case the determinant can be expressed
as an integral over ghost fields, and the gauge-fixed action including
the ghost terms is local, whereas here this is not the case.
This is an important difference, because 
locality is a key ingredient in power-counting arguments, and
thus at the heart of the usual perturbative analysis of
renormalization.

It is therefore of interest to find out whether perturbation theory
can be systematically developed for gauge-fixed Yang-Mills
theory correlated with the soft covariant gauge fixing,
 and its relation to the usual Faddeev-Popov procedure.  
This  question has been addressed some time ago 
in Ref.~\cite{fachin} and recently in Ref.~\cite{golterman,fuji,fuji1,fuji2}. 
A numerical implementation of this technique has been applied
 \cite{parrisoft} to a study of the gluon propagator on small 
lattice volumes but the applicability of this method to
physical lattices seems to be numerically demanding.

\section{The Generic Covariant Gauge}
\label{sec:GENERIC}

In the continuum~\cite{books}, the Faddeev-Popov quantization 
for covariant gauges is obtained by fixing the 
gauge condition 
\be\label{dinamic1}
\partial\m\amu^G(x)=\Lambda(x)\; 
\ee
where $\Lambda(x)$ belongs to the Lie algebra of the
group. 
Since gauge-invariant quantities are not sensitive 
to changes of gauge condition, it is possible to average 
over $\Lambda(x)$ with a Gaussian weight.
As usual the Faddeev-Popov factor can be written as a 
Gaussian integral of local Grassman variables, the resulting 
effective action is invariant under the BRST transformations 
and the correlation functions of the operators satisfy the 
appropriate Slavnov-Taylor identities. The expectation
value of an operator can be cast into 
the following form (to be compared with Eq.~(\ref{eq:omediocont})): 
\be\label{o2}
\langle {\cal O} \rangle =\int\delta\Lambda 
e^{-\frac{1}{ \alpha}\int d^4x Tr(\Lambda^2)}
\int \delta\amu \delta\eta \delta\bar{\eta}\, {\cal O}\, e^{-S(A)-
S_{ghost}(\eta,\bar{\eta},A)}
\delta(\partial\m\amu-\Lambda)\; ,
\ee
obtaining 
\be\label{o3}
\langle {\cal O}\rangle =\int \delta\amu \delta \eta \delta 
\bar{\eta}\, {\cal O}\, e^{-S(A)-
S_{ghost}(\eta,\bar{\eta},A)}
e^{-\frac{1}{\alpha}\int d^4x(\partial_{\mu}A_{\mu})( 
\partial_{\mu}A_{\mu})}\; .\nonumber
\ee
In the perturbative region, the renormalized correlation 
functions can be compared with the same quantities computed 
in the standard perturbation theory.

In Ref.~\cite{noilat} it has been proposed a numerical procedure 
to implement this covariant gauge-fixing on the lattice.
The algorithm is based, as in the Landau case, 
on the minimization of a functional $H_A[G]$ chosen in such 
a way~\cite{giusti} that its absolute minima correspond to a gauge 
transformation $G$ satisfying the general covariant 
gauge-fixing condition in Eq.~(\ref{dinamic1}).
In order to fix the covariant gauge on the lattice,
one should be able to find a functional $H(G)$ stationary when
the gauge (\ref{dinamic1}) is fixed.
The most simple way to define $H(G)$ would be to find a functional 
$h(G)$, to be added to $F$, in such a way that 
$$ \frac{\delta {(F + h)}}{\delta
G}\simeq   \partial_\mu {A^G}_\mu - \Lambda \; . $$
It has been shown~\cite{zwerice,giusti} that this functional 
does not exist for a non abelian gauge theory.
It is interesting to give the outline of the proof.
Writing the gauge transformation:
$$G=\exp{(i \sum_a w^a T^a)}$$
where $T^a$ are the eight $SU(3)$ generators, 
the first derivative of $F$ takes this form
showing the stationarity of $F(G)$ when $\partial_{\mu}A^G_{\mu}=0$:
\be\label{deltaf}
\frac{\delta F(G)}{\delta w^b}=-\frac{2}{g}(\partial_{\mu}
A^G_{\mu})^a\Phi^{ab}(w) 
\ee
where
\be\label{cov10}
\Phi^{ab}(w) \equiv  \left[\frac{e^{\gamma} - I }{\gamma}\right]^{ab} \;
\;\;
\gamma^{ab} \equiv  f^{abc} w^c\;\; .
\ee
Then the derivative of $h$ should have this form:
\be\label{latcov1}
\frac{\delta h}{\delta
w^b(x)}=\frac{2}{w}\left(\Lambda^a(x)\Phi^{ab}(w(x))\right) \;  .
\ee
A necessary condition for the existence of such a 
functional would be
\be
\frac{\delta^2 h}{\delta w^c(x)\delta w^b(y)}= 
\frac{\delta^2 h}{\delta w^b(y)\delta w^c(x)}
\ee
which implies the integrability condition
\be\label{cov3}
\frac{\delta}{\delta w^c(x)}(\Lambda^a(y)\Phi^{ab}(w))=
\frac{\delta}{\delta w^b(y)}(\Lambda^a(x)\Phi^{ac}(w))\; .
\ee
Expanding $\Phi^{ab}(w(x))$ in power of $w(x)$, the Eq.~(\ref{cov3}) 
should be satisfied order by order in $w(x)$. From Eq.~(\ref{cov10}) one 
has
\be
\Phi^{ab}(w)\simeq \delta^{ab} +\frac{\gamma}{2}^{ab}
=\delta^{ab}+f^{abc}\frac{w^c}{2}\; ,
\ee
Equation (\ref{cov3}) is then in contrast with the antisymmetry of $f^{abc}$.\\ 
The new functional,  
chosen to resemble $F$ for this gauge,
 is
~\cite{giusti}: 
\be\label{cov11}
H_A[G]\equiv 
\int d^4x\mbox{\rm Tr}\left[(\partial_{\mu}A^G_{\mu}-\Lambda)  
(\partial_{\nu}A^G_{\nu}-\Lambda)\right]\; , 
\ee
which obviously reaches its absolute minima ($H_A[G]=0$)
when Eq.~(\ref{dinamic1}) is satisfied.
Therefore in this case the Gribov copies
of the Eq.~(\ref{dinamic1}) are associated with
different absolute minima of Eq.~(\ref{cov11}). 
Due to the complexity 
of the functional (\ref{cov11}), it may also have relative 
minima which do not satisfy the gauge condition in 
eq.~(\ref{dinamic1}) (spurious solutions) \cite{giusti}
because the stationary points of $H_A[G]$ actually correspond to the 
following gauge condition 
\be\label{eq:brutta}
D_\nu \partial_\nu(\partial_\mu A^G_\mu-\Lambda) = 0\; ;
\ee  
where $D_\nu$ is the covariant derivative. The spurious solutions, therefore,
correspond to zero modes of the operator
$D_\nu \partial_\nu$.
Of course the numerical minimization of the discretized version of 
eq.~(\ref{cov11}) can reach relative minima (spurious solutions) with
$H_A[G]\rightarrow 0$ which are not distinguishable from the absolute
minima. Hence  this could simulates the effect of 
an enlarged set of numerical Gribov copies.
Preliminary checks at $\alpha=0$~\cite{noilat} do not show any 
practical difference
between the  use of the new functional with respect to the standard
Landau one.

On the lattice, the expectation value of a gauge dependent 
operator ${\cal O}$ in a generic covariant gauge is
\be\label{omedio}
\langle{\cal O}\rangle=\frac{1}{Z}\int d\Lambda
e^{-\frac{1}{\alpha}\sum  Tr(\Lambda^2)} 
\int dU {\cal O}({U^{G_\alpha}})
e^{-\beta S(U)}
\ee
that is the straightforward discretization of Eq.~(\ref{o2})
where $G_\alpha$ is the gauge transformation that 
minimizes the discretized version of the functional 
(\ref{cov11}).
In order to avoid a quadratic dependence on $G$ of  $H_A[G]$ during the 
single, local minimization step of the gauge fixing algorithm,
the discretization
of  $H_A[G]$ has been done by modifying, in each different term 
of $H_U[G]$, the definition of $A$ by terms of order 
a, "driven discretization".
The proposed form of $H_U[G]$ on the lattice is the following: 
\be
H_U[G]  =  \frac{1}{VT a^4 g^2} Tr \sum_{x} J^G(x) 
J^{G \dagger }(x) \; , \label{eq:HJJ}
\ee
where 
\ba
J(x)  & = &  N(x) -i g \Lambda(x)\; , \nonumber\\
N(x)  & = &  -8 I + \sum_{\nu} \left( U_\nu^\dagger(x-\nu) + U_\nu(x)
\right)\; .
\ea
$H_U[G]$ is positive semidefinite and,
unlike the Landau case, it is not invariant 
under global gauge transformations. 
The functional $H_U[G]$ can be 
minimized using the same numerical technique
adopted in the Landau case. In order to study the convergence of the
algorithm, two quantities can be monitored as a function of the number of
iteration steps: $H_U[G]$ itself and
\be\label{eq:theta_Hnew2}
\theta_H = \frac{1}{VT} \sum_x Tr [\Delta_H\Delta_H^\dagger]\; ,
\ee 
where
\ba\label{eq:deltaHnew2}
\Delta_H(x) &  = & \left[ X_H(x) - X_H^{\dagger}(x) \right]_{Traceless}
\propto \frac{\delta H_U[G]}{\delta \epsilon}\; 
\ea 
and 
\ba\label{eq:X_2}
X_H(x) & = &  \sum_\mu \left(U_\mu(x)J(x+\mu) + 
U^\dagger_\mu(x-\mu)J(x-\mu) \right)\nonumber\\ 
& - & 8 J(x) - 72 I 
+ig N(x) \Lambda(x) \; . 
\ea
$\Delta_H$ is the driven discretization of the eq.~(\ref{eq:brutta}) supplemented with
periodic boundary conditions; it is 
proportional to the first derivative of $H_U[G]$ and,
analogously to the continuum, it is invariant under the transformations 
$\Lambda(x) \rightarrow \Lambda(x) + C$, where $C$ is a constant matrix
belonging to the $SU(3)$  algebra. During the minimization process 
$\theta_H$ decreases to zero and $H_U[G]$ becomes constant. 
The quality of the convergence is measured by the final value of $\theta_H$.
 We refer to the original paper~\cite{noilat} for the discussion of 
consistency checks and further details.
In Fig.~\ref{fig:bangaprop} we plot the behaviours of the gluon
propagator tranverse part measured using this technique to fix the gauge
at two different values of the gauge 
parameter $\alpha$~\cite{bangalore}.
A sensitive dependence of the gluon 
propagator transverse part 
on the gauge parameter is clearly reported.
The simulation has been performed over an ensemble of $221$, $\beta=6$
$SU(3)$ 
thermalized configurations with volume $16^3 \cdot 32$.
and has turned out to be moderately time consuming.

Many interesting considerations on the gauge fixing 
related to the gluon propagator can be found in the review~\cite{mandula99}, 
and about the relationship between the gluon propagator and confinement, in 
two recent papers~\cite{cuzaw1,cuzaw2}.

\begin{figure}      
\hspace{2.0cm}  
\ifig{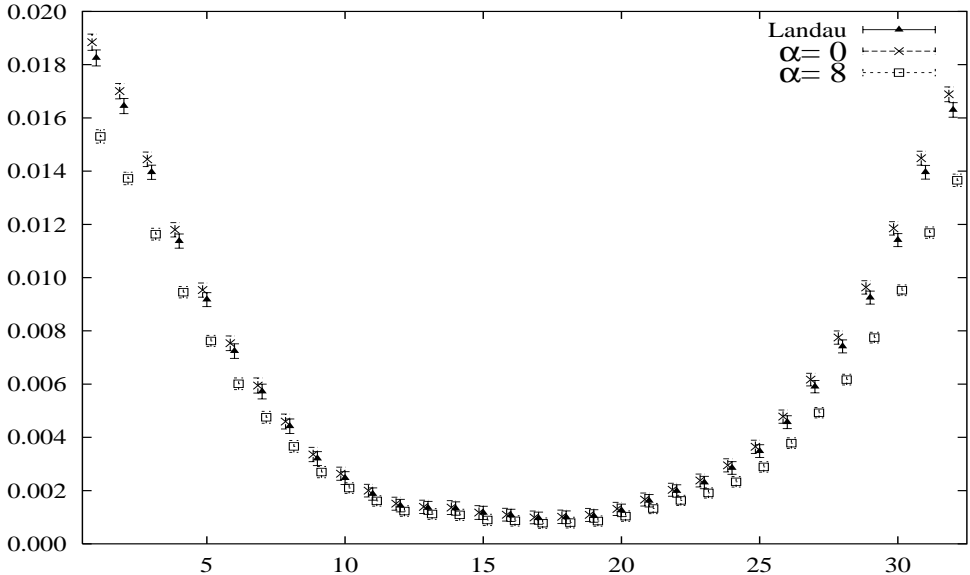}  
  \caption{\small{Comparison of the behaviour of the 
    gluon propagator transverse part as function of lattice time,
   at two different values of the gauge fixing parameter: $\alpha=0$
  (corresponding to the Landau gauge), and $\alpha=8$, for a set of 221,
  $SU(3)$ configurations at $\beta =6$ with volume=$16^3 \cdot 32$.
  From Ref.~\cite{bangalore}.}}
  \label{fig:bangaprop}
\end{figure}
%
\section{The Laplacian  Gauge}
\label{sec:LAPLACIAN}
The Laplacian gauge was proposed in alternative to the standard gauge-fixing 
procedures in order to have a smooth 
gauge-fixing which overcome the problem of Gribov's 
ambiguities \cite{vinkwi92}. The smooth configuration is obtained 
by rotating the gauge
in such a way that the eigenvectors corresponding to the smallest eigenvalues 
of the covariant Laplacian are smooth functions of the lattice coordinates.
The lattice covariant Laplacian is defined as
 \begin{equation}
 \Delta(U)_{ab}(x,y) := \sum_{\mu}[2\delta(x-y)\delta_{ab} - 
     U_{\mu}(x)_{ab}\delta(x + \hat{\mu} - y) - U_{\mu}(x)_{ab}^{\dag}
                    \delta(y + \hat{\mu} - x)]\; , 
  \label{lapla3}
 \end{equation} 
and its eigenfunctions $f^s$ are defined by:	
 \begin{equation}
 \Delta(U)_{ab}(x,y)f_b^s(y) = \lambda^sf_a^s(x)
 \end{equation} 
where $\lambda^s \geq 0$ are the eigenvalues. We have suppressed the gauge 
field indices on $U(x) \in SU(N)$.
The gauge transformation $G(x)$ that defines the Laplacian gauge is computed 
from the $N$ eigenfunctions with the lowest $N$ eigenvalues in order 
to select the smooth modes in the gauge field. 

Specializing to gauge group $SU(2)$ the eigenvalues have a twofold degeneracy, 
due to the charge conjugation symmetry $U = \sigma_2U^*\sigma2$,
 $f^s \to \sigma_2f^{s*}$. The $\sigma_k$ are the usual Pauli 
matrices. The two degenerate eigenfunctions with the smallest 
eigenvalue, $f^0$ and $\sigma_2f^{0*}$, define a $2\times2$ system on all 
sites $x$, which is projected on $SU(2)$ to obtain the gauge 
tranformation $G(x)$,
\begin{equation}
\label{glapla}
{G(x)} = \rho(x)^{-1}i^{1/2}
\left( \begin{array}{cc}
f_1^{0*}(x) & f_2^{0*}(x) \\
if_2^{0}(x) & -if_1^{0}(x) 
\end{array} \right)
\end{equation}
where $\rho(x) = 
(|f_1^{0}(x)|^2 + |f_2^{0}(x)|^2)^{1/2}=1$ 
and the two degenerate eigenfunctions $f^0$ and $\sigma_2f^{0*}$ 
are normalized,
$\sum_x(|f_1^{0}(x)|^2 + |f_2^{0}(x)|^2) = 1$ 
and orthogonal. 

A detailed discussion on the $G(c)$ definition ambiguities rising 
from lowest eigenvalues degeneration and $\rho(x)$'s zero values can be 
found in \cite{vinkwi92}. Nevertheless their effects can be controlled by
increasing the numerical precision with which the lowest eigenvalues and 
eigenfunctions of the Laplacian are computed.

The Laplacian gauge on the lattice is investigated numerically 
in Ref.~\cite{vink94} using the gauge fields $U(1)$ in two 
dimensions and $SU(2)$ in four dimensions. 
The Gribov problem is addressed and to asses 
the smoothness of the gauge field configurations they are 
compared to configurations fixed to the Landau gauge.  
This comparison indicates that Laplacian gauge 
fixing works well in practice and can offer a viable alternative 
to Landau gauge fixing.  
The implementation of this gauge for the $SU(3)$ group has been
studied in Ref.~\cite{frikkand}.
 
A perturbative formulation of the Laplacian gauge for the $SU(2)$ group 
is presented in Ref.~\cite{vanbal94}; however the renormalizability is 
still to be demonstrated. 

\section{The Quasi-Temporal Gauge}
\label{sec:QUASITG}
In  many cases the Landau or Coulomb gauge fixing consume a large
fraction of the computational cost of a simulation.
Therefore it could be extremely advantageous to find 
low-cost alternative gauges with the features of smoothness and
limit to the continuum required by the simulations.
To this aim, in Ref.~\cite{coparpet95} a lattice version of 
the quasi-temporal gauge (QT gauge), proposed
in Ref.~\cite{girotti85} and formulated rigorously in Ref.~\cite{lemiro87}
has been studied.
This gauge is a variation of the temporal gauge, widely studied in 
Ref.~\cite{rossi1,rossi2,rossi3}.
It is defined by enforcing the Coulomb condition at a given
time $t = t_0$:
\begin{equation}
{\Coul}^{\,G}(\vec x,t_0)=0  \;\;\;\;\; \forall (\vec x,t_0).
\label{eq:secondacou}
\end{equation}
 together with the temporal gauge:
\begin{equation}
A_0^{G}(x) = 0 \;\;\;\;\;\;\; \forall x,
\label{eq:tempox}
\end{equation}
at all points.
 The association of the Coulomb gauge on one time slice 
 and the axial gauge makes the quasi-temporal gauge
 a complete gauge with the same properties of the Coulomb one but
 with the advantage of being roughly $T$ times
cheaper to implement on the lattice ($T$ being
 the time direction length of the lattice).
The well-known pathologies of the pure temporal gauge
are overcome by the trick of time slice fixed into the Coulomb gauge. 
In particular the Gauss' law is satisfied and the problematic 
pole of the tree level gluon propagator is removed.
The algorithmical implementation of the quasi-temporal gauge is
straightforward.
Once the Coulomb condition holds at $t=t_0$, the temporal 
gauge~(\ref{eq:tempox}) can be trivially imposed by visiting sequentially 
each timeslice and gauge-transforming the temporal links $U_0 ({\bf x},t)$ 
into the unit group element. 
On a periodic lattice, this can be done for all but one time $t_f$, 
so that the temporal links $U_0({\bf x}, t_f)$, rather than being unity, end up 
carrying the value of the Polyakov loop at ${\bf x}$. 
Since the computational cost of temporal gauge fixing is negligible, 
it follows that the QT gauge is roughly $T$ times faster to implement than the 
Coulomb gauge.
A possible drawback of this gauge condition is that it is not invariant under 
time translation, because of the Coulomb condition at $t_0$.
Moreover, as well as the Coulomb condition, it is affected by the Gribov 
ambiguity. 

In order to test the feasibility of lattice non-perturbative calculations
in the QT gauge, it has been calculated the renormalisation
constant $Z_A$ of the axial current by using WI's on quark 
states. 
This quantity was already  measured with several different 
methods and therefore it is useful to test the quasi-temporal gauge.
The final numerical results of a simulation
on a volume=$V=16^3 \cdot 32$ at $\beta=6$ agrees only roughly with other 
estimates (within errors): the numbers are 
systematically higher than the central value and have larger statistical errors. 
It has been argued that these deviations can be partly related to 
the lattice Gribov ambiguity and that the breaking of 
translational invariance in the time direction may be 
responsible for an enhancement of systematic errors from 
finite volume effects.

\section{Gauge fixing and Confinement}
\label{sec:CONFINEMENT}

In the past decades, many explanations of the QCD confinement 
mechanism have been proposed, most of which share the feature
that topological excitations of the vacuum play a major r\^ole.
Depending on the underlying scenario, the 
excitations giving rise to confinement are thought to be magnetic 
monopoles, instantons, dyons, centre vortices, etc.. 
These pictures include, among others, the dual superconductor picture of 
confinement~\cite{nambu74,parisi75,mandelstam76,thooft76,shilbasch99,digialumo99} 
and the center 
vortex model~\cite{thooft78,amb80,deldebfa97,deldebfa98,lanrete98}. 

Different features of the infrared collective degrees of freedom 
dominating these  two  models  can be identified and 
isolated in different gauges:  the Abelian gauges and the 
Center projection gauges respectively. These procedures end up
with the $SU(N)$ link variables projected as close 
as possible to the elements of $U(N)$ and $Z(N)$ respectively. 
These procedures are explicitly gauge dependent and therefore it is relevant 
to analyze  the projection-physics dependence on the details of the 
gauge fixing procedure.
A brief description of the more favored abelian and center projections, 
will be given in the following.

Reviews on the confinement studies and discussions on the 
implications of the lattice calculations for the question 
whether it is really monopoles or vortices  that drive the 
confining physics or these idea are not necessarily exclusive 
can be found in Refs.~\cite{hay98,hate99,bali99,chegu20,deldig20}

\subsection{Maximal Abelian Gauge}
\label{subs:MAG}
In the scenario of 't Hooft and Mandelstam the QCD vacuum 
state behaves like a magnetic superconductor. A dual Meissner 
effect is believed to be responsible for the formation of thin 
string-like chromo-electric flux tubes between quarks in $SU(N)$ 
Yang-Mills theories. This confinement mecahnism has 
been established indeed in compact $QED$ \cite{vederebashi}. The 
disorder of the related topological objects, magnetic monopoles, 
gives rise to an area law for large Wilson loops and, thus, 
leads to a confining potential.

Nonperturbative investigations of this conjecture became possible 
after formulating the lattice version \cite{kronfeld87} of 
't Hooft's Maximal Abelian Gauge projection   
(MAG) \cite{thooft81}. The idea is to partially fix gauge degrees of 
freedom such that the maximal abelian (Cartan) subgroup ($U^{N-1}(1)$
for $SU(N)$ gauge group) remains unbroken.
Lattice simulations~\cite{kronfeld87, Borny}
have indeed demonstrated MAG to be very suitable 
for investigations of $SU(2)$ abelian projections.
In the case of $SU(2)$ gauge theory, fixing MAG on the lattice amounts 
to maximizing the functional
 \begin{equation}
 F_U[G] = \sum_{x,\mu} Tr \;\left(\sigma_3 U_{\mu}(x)^{G(x)}
                                  \sigma_3 U_{\mu}(x)^{\dagger}{G(x)}\right) 
  \label{mag}
 \end{equation} 
with respect to local gauge transformations $G(x)$.
Condition Eq. (\ref{mag}) fixes  $G(x)$ only up to  multiplications 
$G(x) \to W(x)G(x)$ with $W(x) = exp(\it{i}\alpha(x)\tau_3), \tau_3 = \sigma_3/2, 
-2\pi \leq \alpha(x) < 2\pi$, i.e. $G(x) \in SU(2)/U(1).$
An $SU(2)$ subgroup method~\cite{cama} can be used to perform 
the maximization of the diagonal components of the gauge fields 
with respect to the off-diagonals. The matrix diagonalization  
can be performed iteratively using local gauge 
transformations \cite{okude98} and
overrelaxation can be used in the gauge fixing procedure. 

After that configuration has been transformed to satisfy the 
MAG condition, the coset decomposition:
 \begin{equation}
U_{\mu}(x) = C_{\mu}(x)V_{\mu}(x)
 \end{equation}
is performed, 
where $V_{\mu}(x) = exp(\it{i}\Phi(x)\tau_3), -2\pi \leq \Phi(x) < 2\pi,$
transforms like a (neutral) gauge field and $C_{\mu}(x)$ like a 
charged matter field with respect to transformations within 
the residual abelian subgroup
 \begin{equation}
V_{\mu}(x) \to  W(x)V_{\mu}(x)W^{\dagger} (x+\mu). \; \; 
C_{\mu}(x) \to  W(x)C_{\mu}(x)W^{\dagger} (x+\mu)
 \end{equation} 
Quark fields are also charged with respect to such $U(1)$ 
transformations. The abelian lattice gauge field $V_{\mu}(x)$ 
constitutes an abelian projected configuration.

The $SU(2)$ action of the original gauge theory can be decomposed
into a $U(1)$ pure gauge action, a term describing interactions of the 
$U(1)$ gauge fields with charged fields, i.e. the off-diagonal 
components, and a self-interaction term of those charged fields 
\cite{cherpo95}. 
Maximizing the diagonal components of all gauge fields 
with respect to the off-diagonal components amounts to enhancing 
the effect of the pure $U(1)$ gauge part in comparison with those 
contributions containing interactions with charged fields.
The MAG
projection (and various abelian projections)
might enhance the importance of the  $U(1)$  degrees of 
freedom and the $V_{\mu}(x)$ abelian gauge field 
can be used to investigate: Creutz ratios and Polyakov lines \cite{suyo90}, 
monopoles densities \cite{kronfeld87,ilge99}, dual London relations 
\cite{shilbasch99,sinbro93}, expectation values of monopole creation 
operators \cite{cherpo95,deldebdigia95}, disorder parameters relative to 
monopole condensation \cite{digialumo99}, etc..
\par 
Gribov ambiguities in the actual projection procedure on the 
lattice  will be discussed in Section~\ref{sec:GRIBOV}.
In the continuum, the maximally Abelian gauge,
its defining functional and the Gribov problem (the presence of Gribov 
copies is shown explicitly) in this gauge are reviewed and 
analyzed in depth in Ref.~\cite{bruckhei20}. 

\subsection{Maximal Center Gauge}
\label{subs:MCG}

The old idea about the r\^ole of the center vortices in confinement 
phenomena \cite{thooft78} has been revived recently with the use 
of lattice regularization. In particular, it has been argued that the 
center projection might provide a powerful tool to investigate 
this idea~\cite{deldebfa97}. 
The  gauge dependent studies were done 
in center gauges leaving intact the center group local gauge invariance. 
It is believed that gauge dependent P-vortices (projected vortices) 
defined on the lattice plaquettes are able to locate thick gauge 
invariant center vortices and thus provide the essential 
evidence for the center vortex picture of confinement.
 After gauge fixing the link variables are projected onto the 
centre, i.e. they are replaced by the closest centre element. 
This procedure is in complete analogy with abelian projection 
in the abelian gauges and the center dominance (the 
analog of the abelian dominance) means that the  projected string 
tension $\sigma_{Z(2)}$ ($\sigma_{U(1)}$) is very close to the 
nonabelian theory string tension $\sigma_{SU(2)}$.
Maximal center gauges are defined in the lattice formulation 
of Yang-Mills theory by the requirement to choose link variables 
on the lattice as close to the center elements of the gauge 
group as the gauge freedom will allow. Attempts in this direction 
stemmed from the studies given in Ref.~\cite{deldebfa97}.  
So far three different center 
gauges have been used in pratical computations: 
the direct maximal center gauge \cite{deldebfa98} (described in 
the following), the indirect maximal center gauge \cite{deldebfa97} 
and the laplacian center gauge \cite{deforde99,aledefo99};
the simple center projection gauge fixing procedure 
has furtherly been proposed \cite{scp20} 

The direct maximal center gauge (DMC), 
widely used in $SU(2)$ lattice confinement studies, 
is defined  by the maximization of the functional:
\be  F_U[G] =
 \sum_{x,\mu} \left( \frac{1}{2}Tr U_{\mu}(x)^{G(x)} \right)^2 = 
  \sum_{x,\mu} \frac{1}{4}\left( Tr_{adj}  U_{\mu}(x)^{G(x)} +1 \right),
     \label{maxfunc}
\end{equation} 
with respect to local gauge transformations $G(x)$ (\ref{eq:tras}).
 Condition (\ref{maxfunc})
fixes the gauge up to $Z(2)$ gauge transformation, and can be considered
as the Landau gauge for adjoint representation. Any fixed configuration can
be decomposed into $Z(2)$ and coset parts:  $U_{\mu}(x) = Z_{\mu}(x)
V_{\mu}(x)$, where $Z_{\mu}(x) = \mbox{sign} (Tr U_{\mu}(x))$. The plaquettes
$Z_{\mu\nu}(x)$ constructed from the links $Z_{\mu}(x)$ have values $\pm 1$.
The P-vortices (which  form closed surfaces in 4D space) are made from the
plaquettes, dual to plaquettes with $Z_{\mu\nu}(x) = -1$. The Center 
projection procedure is in complete analogy with Abelian projection in 
the MAG.

The center gauges have been implemented generally for SU(2) \cite{deldebfa97,
deldebfa98,lanrete98,deforde99,fagreole98,enlanre98,chepolves98,bakveszu99,
steph99,langte99}. 
An alghorithm implementing DMC in $SU(N)$ lattice gauge theory 
is proposed and checked on $SU(3)$ vortex-like configurations~\cite{gonzamon98}
 in Ref.\cite{montero20}   

Nevertheless recent alarming results \cite{borko21,borko20} on the 
Gribov problem severity (see Section~\ref{sec:GRIBOV}) in the DMC procedure
cast some doubts on the physical meaning of P-vortices in this gauge.

The continuum analog of the maximum center gauge, in particular of
the Polyakov gauge in which the Polyakov loop has diagonalized, can be 
found in Ref.~\cite{reinengela20}.
\subsection{Issues on string tension}
A matter worthy of note is the measure of the string tension 
extracted from Wilson loops constructed from abelian and center projected 
link variables. 
The observation that the reduced theories reveal the full string 
tension (i.e the abelian or the center dominance) nurtures the 
conjectures that those degrees of freedom give rise to confinement. 
A series of studies \cite{ejkisu} of numerical 
investigations in $SU(2)$ lattice gauge theory has 
established that the abelian projection obtained with 
the MAG fixing indeed accounts for most of the string tension. 
Recent calculations of this quantity for $SU(3)$ can be found 
in Refs.~\cite{wensley20} \cite{yasuki20} \cite{sho20} which 
results are consistent with the values quoted in the 
literature \cite{defor85,mapata95}. In particular in 
Ref.~\cite{sho20} a stochastic gauge fixing method which 
interpolates between the MAG and no gauge fixing is developed. 
The heavy quark potentials derived from Abelian, monopole and 
photon contributions is studied. For Abelian and monopole 
contribution it is observed that the confinement force is 
essentially independent of the gauge parameter.
On the contrary the Gribov ambiguity (see Section~\ref{sec:GRIBOV}) seems to 
influence severely the string tension  in the case of the MCG. 
Recent studies~\cite{borko20,borko21} contradict 
previous  numerical simulations  demonstrating that the 
entire asymptotic string tension was due to vortex-induced fluctuations 
of the Wilson loop~\cite{koto99}.

\section{Gauge Fixing in the Langevin Scheme}
\label{sec:LANGE}
In this Section we will discuss the use of gauge fixing 
in algorithms in which the field configurations
are generated using a discretized Langevin equation.
This numerical technique has been adopted to implement the 
so-called difference method~\cite{Paris81,Falcio83,Falcio85} 
whose relevance, both for gauge and spin systems, has been 
widely acknowledged (see for example Ref.~\cite{mapata95}). 

In the Monte Carlo approach, correlation functions are obtained 
as  expectation values of operators over a suitable 
number of uncorrelated configurations. 
As they are generally expressed as differences between similar
numbers, they can be affected by large statistical fluctuations. 
The difference method  is an interesting attempt
to override this effect to get more accurate results.
This method is based on the idea of perturbing the system far 
away from equilibrium in a limitated space-time region and 
measuring the decay of the correlations from such zone by 
using specific properties of the dynamical updating algorithm.
In this procedure one computes the differences between  the 
matrix element values evaluated on a perturbed configuration
and on the unperturbed ones. If one is able to keep the two 
configurations very close each other, a coherent cancellation of 
statistical errors between the two highly correlated stochastic 
processes follows in the difference. 
To describe the difference method let us consider two gauge systems
$K$ and $K'(i)$ 
on two lattices of identical size $M^3 \times L $; $K$ is kept at 
inverse square coupling $\beta$; $K'(i)$ has a  ``time slice'' $i$ 
$(i = 1,2,3 ... L)$ where the inverse square coupling takes the 
value $\beta +\delta\beta$. 
We define $E(j)$ as the  average energy of the time slice $j$ for 
the system $K$, and $W(i,j)$ as the average energy of time slice 
$j$ for the perturbed system $K'(i)$. Then it is possible to  
show \cite{Falcio83} that the connected  
correlation function at distance {\em d} of the energy operator
$C(d)$ ( using Wilson action) is
\begin{equation}
  C(d) = \frac {E(j) - W(i,j)}{\delta\beta}\ ,
\end{equation} 
for any $i,j$ such that $d = \vert j - i \vert$. 
In order to  avoid noisy correlations the two sets of 
configurations   must 
be {\em similar} (we are forgetting for a moment about gauge
invariance) and this happens only if the method used to generate the
configurations is continuous in the $\beta$ variable.

The Langevin \cite{Halp77,Paris80,Paris81,Falcio83,
Hamber84} update scheme is a well known example of continuous 
algorithm. Here we will give some details of this scheme; it 
will be useful also in the following to define the numerical 
stochastic perturbation theory.
If $S$ is the action, the gauge field configurations can 
be obtained as solution of the following  Langevin equation:
\begin{equation}
  \dot{U}_L(t) = \frac{\delta S}{\delta U_L} + \eta_L(t)\ ,
  \label{eq:langeq}
\end{equation} 
where $\eta_L(t)$ is a gaussian noise with autocorrelation:
\begin{equation}
  \langle \eta_L(t)\eta_{L'}(t') \rangle = \delta_{LL'}\delta (t-t')\ .
\end{equation}
Notice that the ``time'' $t$ occurring in Eq.~(\ref{eq:langeq})
has nothing to do with the physical euclidean time.
In fact, it is the evolution time 
 of the differential equation  dynamics adopted 
to formulate Langevin algorithm in lattice gauge theory. 
After a certain lapse of time, the system 
will become representative of the Boltzmann distribution; 
therefore the Langevin equation provides a way to generate 
this distribution analytically and numerically.

Some extra care is needed to discretize the Langevin equation 
as new parameters are needed; they have to be tuned in order to
get good performances.
Just to be concrete, we sketch here the implementation of the 
Langevin dynamics suitable for lattice simulations, from 
Ref.~\cite{bakako85}. A single Langevin step is given by a 
sweep of the lattice where each link variable is updated 
according the rule
\begin{equation}\label{latlange}
U_\mu(x) \to U_\mu'(x) = e^{-F_\mu(x)} \; U_\mu(x)
\end{equation}
where $F_\mu(x) $ is given by
\begin{equation}
F_\mu(x) ={\epsilon\beta\over 4 N} \sum_{P\supset \mu}(U_P - U_P^\dagger)
\vert_{traceless} + \sqrt\epsilon H_\mu(x)
\end{equation}
Here $\epsilon$ is the Langevin time step; 
the sum over $P$ means that $F_\mu$ gets contributions from
all {\sl oriented} plaquettes which include the link $\mu$ at $x$.
Finally $H_\mu(x)$ is extracted from a standard
(antihermitian, traceless) Gaussian matrices ensemble. 

\subsection{The Chameleon Gauge}

The effectiveness of the difference method implemented by the 
Langevin dynamics depends on the evolution trajectories in the 
phase space of the two systems, unperturbed and perturbed: 
they should be as close as possible. 
For a gauge model, the additional degrees of freedom 
create an additional complication. The gauge part 
performs a random walk in phase space, and 
tends to separate the two trajectories of the Langevin 
dynamics~\cite{Paris81, Falcio83}.
A simple gauge fixing (for example putting all the time-like
gauge variables = 1) 
turns out to slow down the dynamics and makes the method
impractical~\cite{Falcio83}. Also the introduction 
in the Langevin equation of a magnetic 
field term \cite{Falcio83}, which would ensure a 
smooth, partial gauge fixing, does not turn out to be 
successful. Trajectories in phase space diverge quite 
soon and there is a very unpleasant  slow drift of the energy. 

To overcome these problems a peculiar gauge fixing called 
{\it chameleon gauge} was proposed~\cite{Chameleon}.
 The gauge is fixed 
in such a way that two fields are as similar as possible. 
This procedure does indeed 
keep the gauge part of two systems as close as possible reducing 
the rate of divergence of the two trajectories in phase space 
without introducing any sizable slowing down in the 
 observable  dynamics like the energy.
After each full lattice sweep of the Langevin updates of link 
variables $U$ (unperturbed fields) and $V$ (perturbed 
fields), a gauge fixing is performed on the $U$ configuration. 
One  maximizes  the quantity:
\be
 F_U[G] = \sum_{x,\mu}U_{\mu}^G(x)V_{\mu}^{\dagger}(x)
\ee
where $x$ runs over the lattice sites and $\mu$ over the 4 
directions, with respect to gauge transformations $G(x)$. The 
gauge fixing performed at each step has a sizable 
effect on the evolution of the system as the corresponding 
Eq.~(\ref{eq:langeq}) is not gauge invariant.
The quantity
\be
{\cal F} \equiv \sum_{x,\mu}\left(1-U_{\mu}(x)V_{\mu}^{\dagger}(x)
\right)
\ee
measures the gauge fixing quality, small values of $\cal F$ 
mean good gauge fixing. 
Correlations functions for the 
$SU(2)$ $0^{++}$ glueball mass measured on the chameleon 
gauged configurations are far less noisy (a factor of order 5) 
than with other methods. It is also quite interesting to note 
that the breakdown of the correlation functions is always 
signaled by a sudden growth of the quantity, i.e. by the 
collapse of the quality of the gauge fixing one is able to 
reach.
In Fig.~\ref{cham1} it is shown the signal extracted at separation 
1 by using the magnetic field method \cite{Falcio83}, while in 
Fig.~\ref{cham2} it is shown the result obtained with chameleon 
gauge fixing.  
\begin{figure}      
\ifig{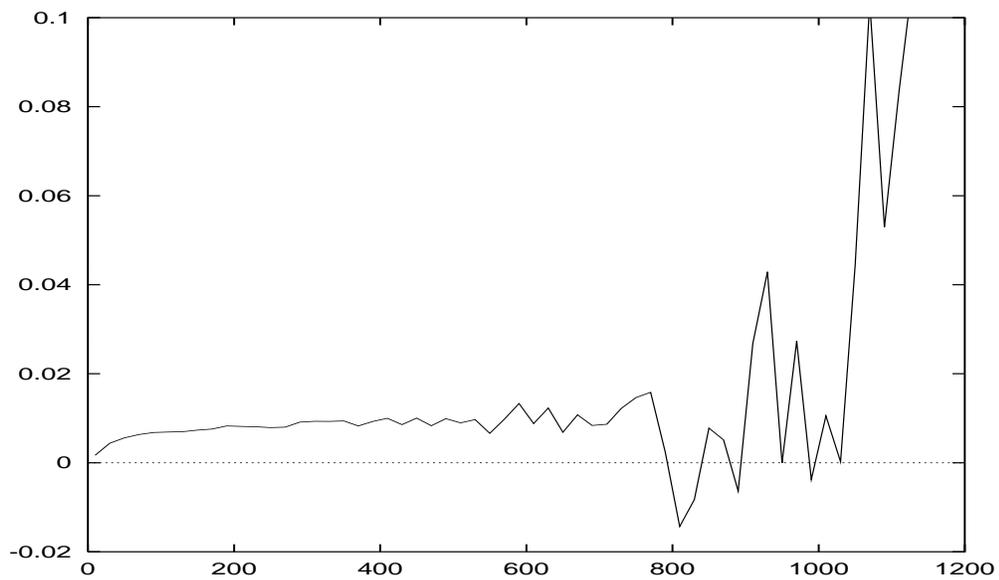}  
 \caption{The distance 1 correlation computed by using the magnetic 
field method.} 
 \label{cham1}
\end{figure}
\begin{figure}      
 \ifig{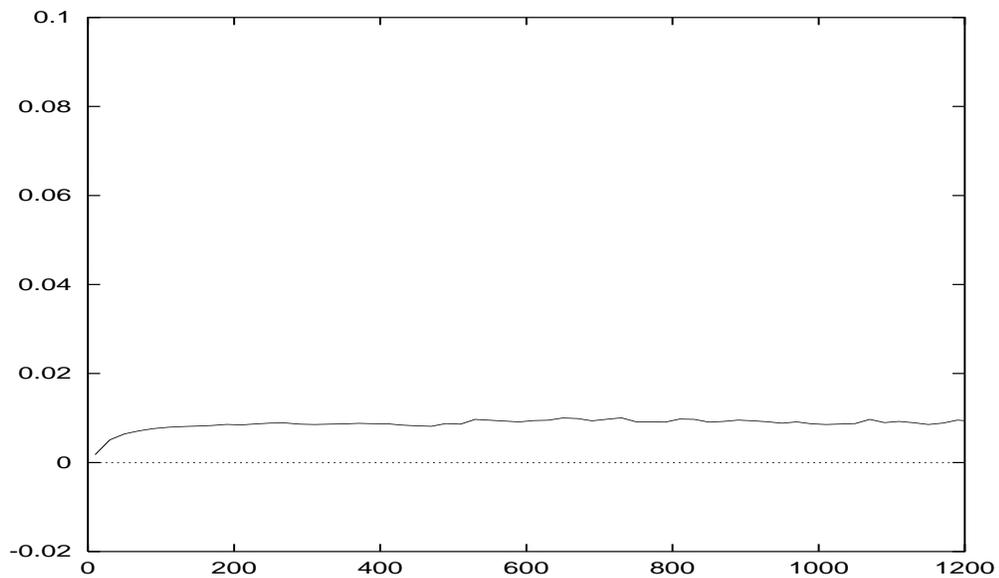}  
 \caption{The distance 1 correlation computed by using the chameleon 
gauge fixing.} 
 \label{cham2}
\end{figure}

\subsection{Numerical Stochastic Perturbation Theory}
To compute large orders in the 
perturbative expansion of observables (therefore gauge invariant) 
in lattice field theories and also to the aim of 
overcoming some divergent 
fluctuations occurring in perturbative Langevin dynamics, 
the method of the numerical stochastic perturbation theory 
(NSPT) has been proposed in 
Ref.~\cite{direnzo0,direnzo2}.
This method is implemented in the scheme of stochastic gauge fixing 
originally formulated  in the continuum \cite{zwan81} 
and then extended to the lattice case \cite{rodale88}. 

In this approach, perturbation theory is performed through 
a formal substitution  of the expansion ($k$ is the perturbative 
order and $g$ is the standard coupling in lattice gauge 
theories):
\be
U_{\mu}(x)  \to  \sum_{k}g^kU_{\mu}^{(k)}(x)
\ee   
in the  Langevin equation (\ref{latlange}) used for lattice 
simulations. 
The power expansion of the field induces a power expansion
 for every observable ($A_{\mu}(x)$ included), then
the 
perturbative expansions is usually computed as an average over 
the Langevin evolution. 
Even though the original motivation of the Langevin approach was 
to perform calculations in perturbation theory without 
fixing a gauge, it is known that some divergent fluctuations 
 may plague high order terms (averaging to zero). A proposed way out 
is the technique of the stochastic gauge fixing. 
The underlying idea of this approach is the introduction of 
an attractive force in the Langevin equation in such a way 
that the field is attracted by the manifold defined by Landau 
gauge and that its norms are kept under control without affecting the 
observables.
The implementation on the lattice consists in a gauge 
transformation, which is executed after each Langevin step, 
given by
\begin{eqnarray}
{\cal W}_L: U_\mu(x) &\to& e^{w(x)} U_\mu(x) e^{-w(x+e_\mu)}\nonumber \\
w(x) &=& \alpha\; \sum_\mu\;\Delta_{-\mu}
[U_\mu(x)-U^\dagger_\mu(x)] \Vert_{traceless}\nonumber \\
\Delta_{-\mu}U_\nu(x) &\equiv& U_\nu(x)-U_\nu(x-e_\mu)
\end{eqnarray}
One can prove \cite{direnzo3} that by doing this, the system 
gains a force that drives it towards the Landau gauge. By 
interleaving it to each Langevin step one obtains a sort of 
soft gauge fixing where  ${\cal W}_L$ provides an additional drift 
which however does not modify the asymptotic probability distribution.
After that, one has  to expand in $g$ the gauge fixing step ${\cal W}_L$:
 this can  be achieved with 
the same technique already developed for the unconstrained
Langevin algorithm \cite{bakako85}. The value of the
 parameter $\alpha$ is 
chosen in such a way to minimize systematic errors. 
We report, for example, the term in $g^4$ of the plaquette 
measured in Fig.~\ref{plaq1} without gauge fixing and 
in Fig.~\ref{plaq2} with the above gauge fixing.
\begin{figure}      
\ifig{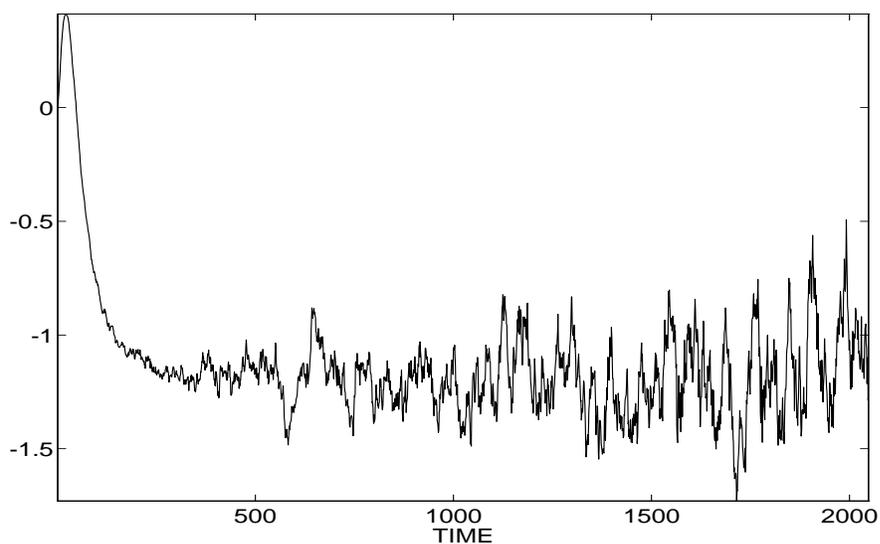}  
 \caption{The fourth order in the expansion of 
  the plaquette in a pure Langevin simulation.} 
 \label{plaq1}
\end{figure}
\begin{figure}      
 \ifig{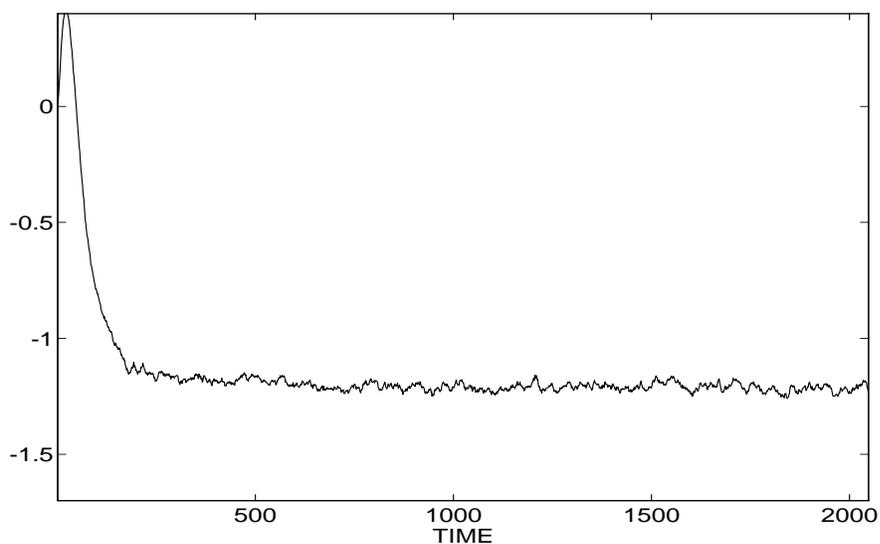}  
 \caption{The fourth order in the expansion of 
  the plaquette in a  simulation with stochastic gauge fixing.} 
 \label{plaq2}
\end{figure}                     
The extension of NSPT with the adoption of the stochastic gauge 
fixing  in a gauge non invariant context is possible with the 
caveat that also the gauge condition one wants to enforce has to 
be expanded as a series of conditions. To be definite, as the 
$A_{\mu}$ field is expanded as 
$A_{\mu} = A_{\mu}^{(0)} + gA_{\mu}^{(1)} + g^2A_{\mu}^{(2)} + ...$, 
the form of Landau conditions one needs to impose is 
\be
\partial_{\mu}A_{\mu}^{(k)}  = 0 
\label{landpert}
\ee
for every order $k$ (partial derivatives, as usual, are to be 
understood as finite difference operators). The implementation~
\cite{direnzo1} goes as follows.
The functional, as function of $A_{\mu}$, to be maximized by the gauge 
transformation, is: 
\begin{eqnarray}
{F_A[W]} &=& \sum_{x,\mu}Tr\left(A_{\mu}^WA_{\mu}^{W\dag}(x)\right)\nonumber \\ 
       &=& \sum_kg^kN^{W(k)}(x) 
\end{eqnarray}
where an expansion for $F_A[W]$ is induced by the expansion of the 
field $A_{\mu}$.
If the transformation $W$ is chosen as 
\begin{eqnarray}
W(x) &=& \sum_{k\ge0}g^kw^{(k)}(x) \\
w^{(k)}(x) &=& -\alpha\sum_{\mu}g^k\partial_{\mu}A_{\mu}^{(k)(x)}\,,\;\;
\; (0 <\alpha < 1)  
\end{eqnarray}
then the extremum conditions for every $N^{(k)}(x)$ are recovered 
enforcing 
exactly eq.~(\ref{landpert}).

The strategy of the  extension of  NSPT to compute 
the expansion of gauge non invariant quantities~\cite{direnzo3,direnzo1}, 
is the following:
\begin{itemize}
\item[i)] let the system evolve (in the stochastic gauge fixing 
scheme) to get a thermalized  configuration;  
\item[ii)] fix Landau gauge implementing  the condition~(\ref{landpert}) 
order by order and measure;
\item[iii)] go back to i), i.e. let the system evolve until a decorrelated 
configuration is reached.
\end{itemize}

The status of the method with respect to gauge fixed 
lattice $QCD$ is revised and a first application to compact 
(scalar) $QED$ is presented in Ref.~\cite{direnzo3}. 
A discussion about the convergence of the stochastic process towards the
equilibrium, the expected 
fluctuations in the observables and the computations of 
quantities at a fixed (Landau) gauge in this frame can be 
found in Ref.~\cite{direnzo4}. A success of the NSPT extension  
has been the computation of the lattice $SU(3)$ basic 
plaquette to order $\beta^{-8}$\cite{direnzo2} to actually verify 
the expected dominance of the leading infrared (IR) renormalon 
(a recent review on this item is in Ref.\cite{beneke}) 
associated to a dimension four condensate. Recently the order 
$\beta^{-10}$ has been computated~\cite{diresco} and then result is 
consistent both with the expected renormalon behaviour and with 
finite size effects on top of that. Another application of
the NSPT method has been the computation of the perturbative expansion of 
the so called residual mass term in lattice heavy quark 
effective theory to order $\alpha_0^3$\cite{diresco1}.

%

  \section{Lattice Gauge Potential}\label{sec:LGP} 

In this Section we will discuss the problems related to 
the ambiguities due to the lattice definition 
of the gauge potential $A_\mu$. 
The root of the problem is that a unique, natural definition of the 
potential $A_{\mu}$ on the lattice does not exist because
in the Wilson discretization of gauge theories, the fundamental fields
are the links $U_\mu$ which act as parallel transporters of the theory.
Hence the lattice fields $A_\mu$  are derived quantities which 
tend to the continuum gluon field as the lattice spacing 
vanishes. As a consequence on the lattice it is possible to choose different
definitions of $A_\mu$ formally equal up $O(a)$ terms and there is not
any theoretical reason to prefer one or another.
In quantum field theory this ambiguity is well understood because 
any pair of operators differing from
each other by irrelevant terms, i.e. formally equal up to terms of order 
$a$, will tend to the same continuum 
operator, up to a constant, see for a general discussion Ref.~\cite{testabs}. 
In the Wilson's regularization the natural and most used definition of the 
4-potential in terms of the links, $U_{\mu}$, which represent
the fundamental dynamical gluon variables is given by 
the standard relation (\ref{eq:standard}).
This definition is certainly not the only possible and this 
ambiguity can create some problems in the discretizations
of the continuum gauge fixing equations.
Other definitions with analogue properties as, for instance
\be
A^{\prime}_{\mu}(x)
\equiv  \left[{(U_{\mu} (x))^2 - (U_{\mu}^{\dagger)} (x))^2\over
{4 i a g}}\right]_{Traceless}\;
\label{eq:quarta} 
\ee
 which in fact differs from to the standard one (\ref{eq:standard}) 
by terms of $O(a)$ that formally go to zero as $a \to 0$, have the
same validity.

Of course the requirement that
the gluon fields $A_{\mu} (x)$ rotated in the Landau or Coulomb gauge
satisfy the corresponding gauge conditions (see Section~\ref{sec:LGF})
is based on the gluon field definition, therefore different $A_{\mu}$ 
definitions generate different gauge fixings on the lattice which
in the limit
$a \to 0$ must correspond to the  
continuum gauge fixing.
This feature, checked in perturbation theory,
has been verified numerically at the non-perturbative level 
in Ref.~\cite{giusti1}, where it has been shown that different 
definitions of the gluon field give rise to Green's functions
proportional to each other, guaranteeing the uniqueness of the 
continuum gluon field.

 The relation between two $A_\mu$ definitions can be expressed up to
 $O(a^2)$ terms in this way \cite{testabs}:
\be
{A^{'}}_{\mu}(x)=C(g_0) A_{\mu}(x). \label{eq:relazione}
\ee
Therefore for a Green's functions insertions the following
ratio is expected to be a constant
\be
{{\langle \dots A'_{\mu}(x) \dots \rangle} \over {\langle \dots A_{\mu}(x)
\dots \rangle}}= C(g_0) \; . \label{green}
\ee
This relation has been checked numerically on the lattice
by measuring 
a set of Green's functions related to the gluon propagator 
for $SU(3)$ in the Landau gauge with periodic
boundary conditions computed
with the insertion of the standard $A_\mu$~(\ref{eq:standard}) 
and $A^{\prime}_\mu$~(\ref{eq:relazione}).
\begin{figure}
\bc
\ifig{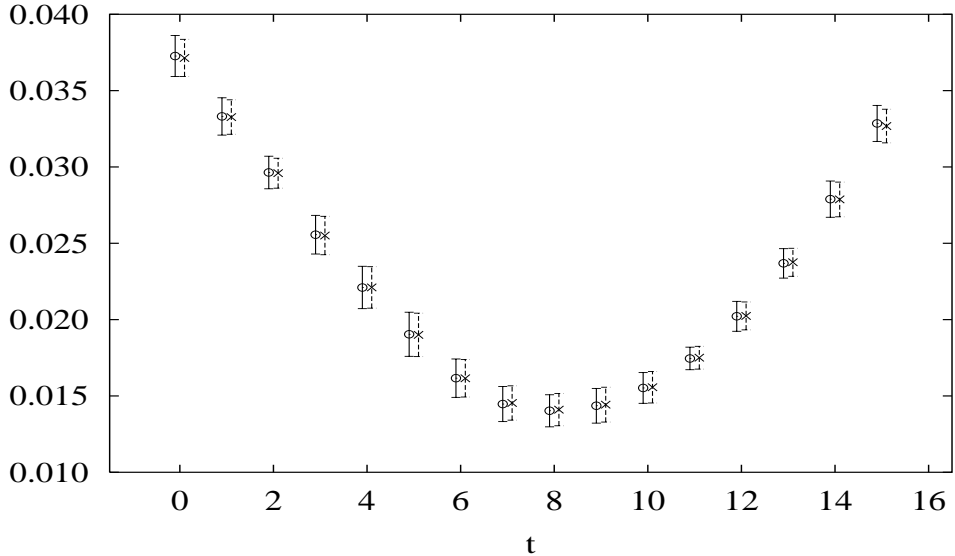}
\caption{\small{Comparison of the matrix elements of  $\langle{\AC}^{'}_i{\AC}^{'}_i\rangle(t)$ 
(crosses) and the rescaled $\langle{\AC}_i{\AC}_i\rangle~\cdot~C_i^2(g_0)$
(open circles) as function of time for 
a set of 50 thermalized $SU(3)$  configurations at $\beta=6.0$ with a 
volume $V\cdot T=8^3\cdot 16$. The data have been
slightly displaced in $t$ for clarity, the errors are jacknife.
From Ref.~\cite{giusti1}.}}
\label{fig:amu}
\ec
\end{figure}
In Fig.~(\ref{fig:amu}) $\langle {\AC}^{'}_i{\AC}^{'}_i\rangle$
and the rescaled one
$C_i^2(g_0) \langle {\AC}_i{\AC}_i\rangle$ are shown, where
\be
\langle {\AC}_i{\AC}_i\rangle (t) \equiv  \frac{1}{3 V^2}  
 \sum_{i}\sum_{{\bf x},{\bf y}} Tr \langle  A_i({\bf x},t)A_i({\bf y},0)\rangle 
\label{eq:AiAi}\\
\ee
and the matrix elements $\langle {\AC}^{'}_i{\AC}^{'}_i\rangle (t)$
are obtained replacing in the same form the alternative 
definition~(\ref{eq:relazione}).
The remarkable agreement between these two quantities
confirms the proportionality shown in Eq.~(\ref{eq:relazione}).

From the algorithmical point of view, however, the 
various definitions are not interchangeable. 
In fact  the parameter $\theta$, 
monitoring the numerical 
behaviour of the gauge fixing algorithm (as explicated 
in Section~\ref{sec:LGF}), as a function of lattice sweeps,
is quite sensitive to different $A_{\mu}$ definitions.
As shown in Fig. (\ref{fig:p60t}) only the $A_{\mu}$ definition which
appears in the $F$ functional  minimization, see Eq.~(\ref{eq:landlat}),
goes to zero.
\begin{figure}      
\hspace{2.0cm}  
\ifig{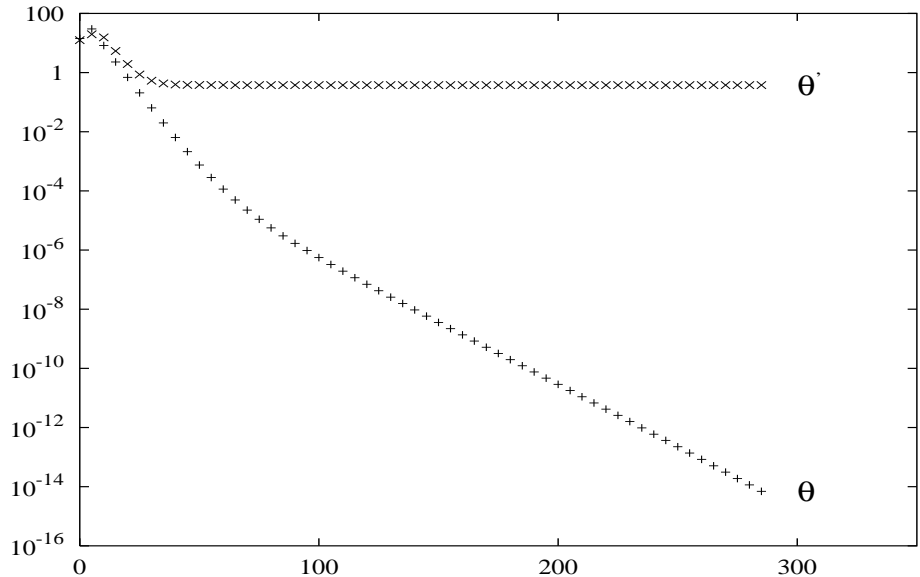}  
  \caption{\small{Typical behaviour of $\theta$ and $\theta^{'}$ vs gauge 
  fixing sweeps at $\beta=6.0$ for a thermalized $SU(3)$ configuration 
  $8^3\cdot 16$} from Ref.~\cite{giusti1}.}
  \label{fig:p60t}
\end{figure}
Note that the comparison reported in Fig.~\ref{fig:p60t} is done on one
configuration at time in order to
check the gauge fixing  quality while the comparison between 
the operators $\theta$ and $\theta^{\prime}$
must be done averaging them
over the gauge fixed configurations of the thermalized set.
Moreover the behavior shown in Fig.~(\ref{fig:p60t}) can be readily understood
in the following way.
The operator $\theta $, defined in Eq.~(\ref{eq:thetalat}),
and the operator 
 ($\theta^{'}$), constructed with the same form with $A_\mu$ 
 replaced by ${A^\prime}_\mu$, 
 are computed in the lattice units taking the definition  
eq.(\ref{eq:standard}) and  Eq.~(\ref{eq:quarta})
 without the powers of $a$ to the denominator.
Then in the continuum variables
$\theta =\frac{a^4}{\VC}\int d^4x (\partial_\mu A_\mu(x))^2$
 where $\VC$ is the 4-volume in physical
units  and analogously  for  $\theta^{'}$.
Hence, while $\theta$ vanishes configuration by configuration,
as a consequence of the gauge fixing, $\theta^{'}$ is proportional
to $(\partial_\mu A'_\mu)^2$, which has the vacuum quantum numbers and
mixes with the identity. The expectation value of
$(\partial_\mu A'_\mu)^2$, therefore, diverges as ${1 \over a^4}$
so that $\theta^{'}$ will stay finite, as $a \rightarrow 0$.

As a matter of fact, the construction of lattice operators 
converging, as $a \to 0$, to the fundamental continuum gauge fields, is 
affected, at the regularized lattice level, by an enormous 
redundancy due to the irrelevant terms. 
Although, if on the general field theoretical grounds the 
validity of such results is not unexpected, it is
remarkable that it holds  true also in this particular 
situation in which gauge fixing is naively performed, 
disregarding the problems related to the existence of 
lattice and continuum Gribov copies (see Section~\ref{sec:GRIBOV}).

The freedom to choose the lattice definition of 
$A_\mu$ can be used to build discretized functionals
which lead to more efficient gauge-fixing algorithms. 
In Ref.~\cite{bonnet20} a new gauge fixing functional is proposed 
to remove the lattice discretization 
errors  of order $ O(a^2)$ to the Landau gauge condition. This 
improved scheme is used to fix the Landau gauge in the $SU(2)$ lattice 
simulation of the  gluon propagator in Ref.~\cite{chenhe20}.

A Landau gauge fixing algorithm, using the exponential relation
between link and gauge field
is studied in Ref.~\cite{naka99}. 
In Ref.~\cite{cucchie99} the gluon propagator is evaluated in different 
gauges and using different gluon field definitions on the lattice, 
corresponding to discretization errors of different orders.

  \section{Lattice Gribov Copies}
\label{sec:GRIBOV}
In 1978 Gribov~\cite{Gribov} discovered that for non-abelian 
gauge theories the usual 
linear gauge conditions does not fix in
a unique way the gauge potential. 
In fact, it is possible to find different 
gauge potentials satisfying the gauge condition which are
related each other by nontrivial gauge transformations. 

The presence of Gribov copies implies that 
the constraint of the na{\"\i}ve gauge fixing
is not sufficient to remove all the degrees of 
freedom associated to the group of gauge transformations.

It is interesting to discuss the gauge fixing procedure
 in  geometrical language, see for example Ref.~\cite{BAAL}. 
The point where the gauge orbit, the curve described by field $A^{G}$
as a function of the gauge transformation $G$, intercepts the plane defined by 
the gauge condition $f(A^{G})=0$ 
represents a gauge-fixed potential. If the gauge orbit intersects 
in more than one point the plane $f(A^{G})=0$, each
point represents a Gribov copy.
It is possible to define a Hilbert norm of the gauge potential along the orbit
as given in Eq.~(\ref{eq:effe}). This definition has all the good properties of 
a norm and its values are
able to distinguish among different local gauge transformations.
Global gauge transformations does not change the norm
and therefore they have to be considered in the same
class of equivalence.
The gauge transformation(s) capable to enforce the gauge condition can be
found searching the stationary point(s) of the following
functional (see also Eq.~(\ref{eq:effe})):
\begin{equation} \label{eq:Fnorm}
F_A(G)=-||A^G||^2=-\sum_\mu \int d^4x Tr[(A{^G}_\mu (x))^2] \;.
\ee
In the case of minima of $F$ the determinant of the Hessian matrix 
of the second derivatives of $F$ is positive,
this coincides with the Faddeev-Popov determinant (in the case of the Landau
gauge, for example, the Faddeev-Popov operator $FP$ is $FP=-\partial D[A^G]$).
The set of the gauge potentials $A^G$ which are the minima of the corresponding
$F$'s defines the Gribov region $\Omega$ which is known to be a convex
region and its boundary $\partial \Omega$,  where
lowest eigenvalue of the Faddeev-Popov operator vanishes,
is known as  the (first) Gribov horizon. Note that the minima of $F$
satisfy a more restricted gauge condition than the simple Landau gauge fixing
which requires only the stationarity of $F$.
This  restriction,
unfortunately, is not sufficient
to  solve the problem of Gribov copies because it is known that
there can be multiple intersections also inside the Gribov region.
The solution can be found by requiring a more restricted region of absolute
minima, called the {\it fundamental modular domain}, contained inside the
Gribov region. 
This scenario shows a possible attempt to solve the Gribov problem based on 
very interesting properties of the gauge potential topology.
Unfortunately it is impossible to implement this constraint numerically.

The studies of the Gribov problem on the lattice have 
demonstrated the existence of gauge fixing ambiguities 
both for abelian and non abelian theories for the first time 
in Refs.\cite{nakasi90,forcra91,em91} in the case 
of Coulomb and Landau gauges.
It is remarkable
that the scenario on the lattice seems to follow very closely
the continuum one. In fact the lattice Gribov copies are found to correspond
to different minima of the functional $F$ so that they appear inside
the lattice Gribov region. The solution could be to take the absolute
minimum, which certainly exists in the compact theory, but this
prescription is numerically hopeless.
It must be noted also the corrispondence between the procedure
and the formulas used on the lattice and in the continuum.
Nevertheless, even if the phenomenological situation seems 
to be equivalent to the continuum one, it must be clearly 
understood that the Gribov copies are, at most or likely 
almost completely, determined by numerical lattice artefacts.
On the other hand, 
in the continuum case the reason for the presence of  Gribov 
copies is related to the deeper level of the theory, being 
usually connected with topological obstructions which forbid
the possibility to set up a smooth, diffential gauge condition 
valid over all the lattice.
Moreover, it is very difficult to make any connections between
lattice Gribov copies and the continuum ones because the study of
the continuum limit is operatively based on a chain of lattice 
simulations at different $\beta$ values whose gauges change without 
any possibility to be controlled.
Therefore the lattice studies of the Gribov ambiguity can have only 
a moderate influence on the theoretical uncertainties.
Nevertheless the lattice Gribov copies may have a r\^ole in numerical
simulations which therefore, must be kept under control.

The generation of Gribov copies on the lattice 
is by now a standard method. Here we describe the procedure
(mother and daughter method) which is schematically 
described in Fig.~\ref{F_madre}.
\begin{figure}      
\hspace{2.0cm}
\includegraphics[height=160mm,width=130mm]{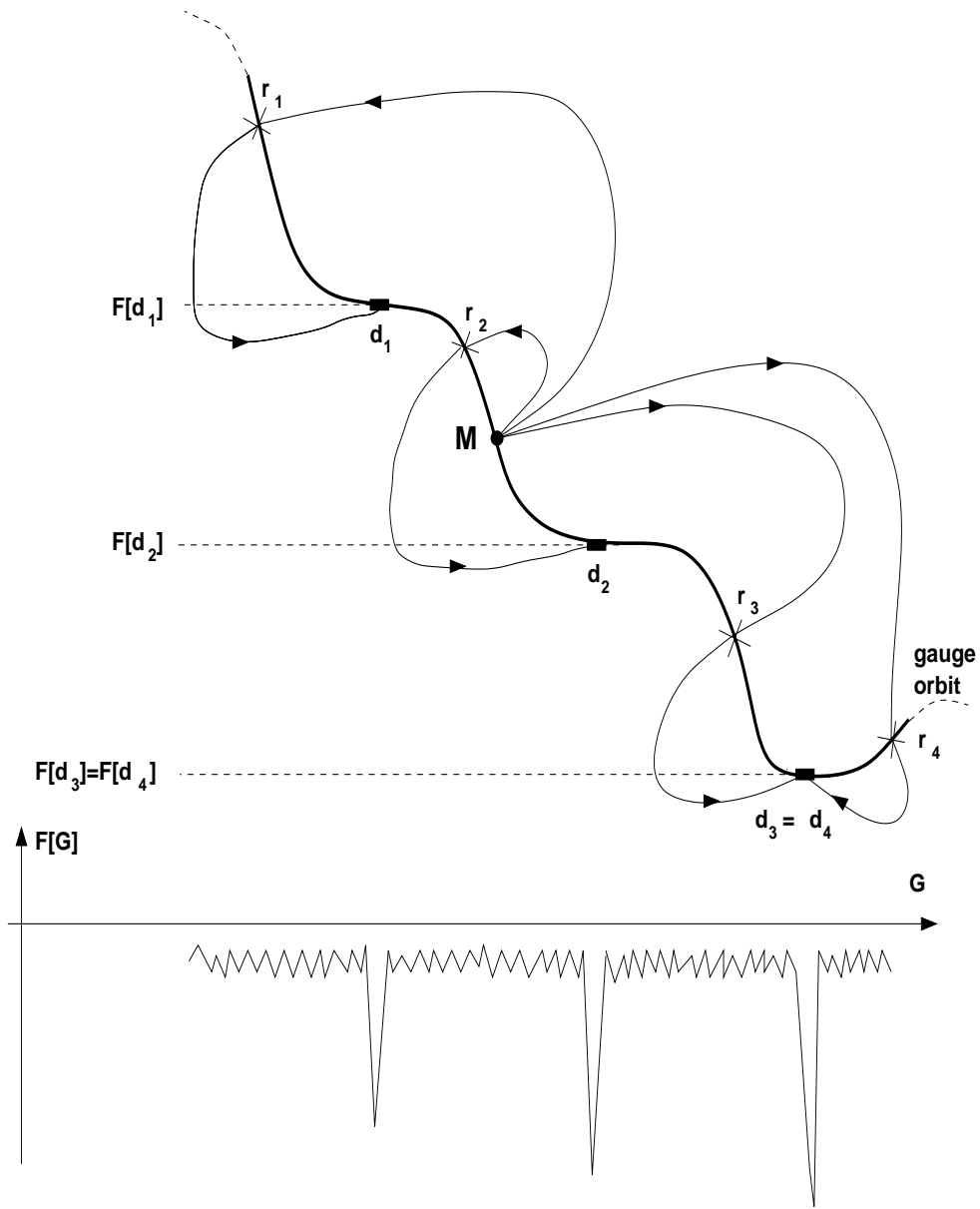}
 \caption{Schematic cartoon describing  the usual method to generate
 lattice Gribov copies by means of random gauge rotation.} 
 \label{F_madre}
\end{figure}
Given a thermalized  
configuration {\bf M}=$\{U\}$, where {\bf M} is for mother, 
generated by a Monte Carlo simulation, 
one applies on it an ensemble of random, local gauge transformations.
This is the cheap part of the procedure because the generation of
random gauge transformations and the gauge rotations require a very short
computer time.  
In such a way an ensemble 
of configurations ${{\bf r}_i}$ gauge related to the mother are generated.
Then, and this is quite expensive part in computer time,
 all these 
configurations are gauge fixed obtaining, at the end, the ensemble
of the daughters ${{\bf d}_i}$.
The final step consists of the analysis of the gauge-fixed 
ensemble, in order to understand whether all the members 
of the ensemble 
have been fixed
to the same gauge configuration, or 
different minima  (i.e. lattice Gribov copies) have appeared.
In order to perform such a test, a good quantity 
to be measured is  
the final value of the functional $F [G]$ defined in 
Eq.~(\ref{eq:effel}). In fact the $F$ value 
is naturally gauge dependent and it is not affected by 
 global gauge transformations
which are to be considered gauge-equivalent, i.e.
related to each other by global gauge transformations $G(x)=G$.
Of course other gauge dependent forms ~\cite{davi,em91,
parpetvla91} may be adopted.

To summarize: on the lattice  the existence of many different 
minima of the functional $F$, not gauge equivalent, are 
called lattice Gribov copies and can be labelled with the 
value of the functional $F$ itself.
Of course it is unthinkable to succeed in reaching numerically 
the absolute minimum. The search of the  $F$ minima is at least 
as difficult as to find the lowest state of energy of a spin 
glass system with  hamiltonian $F$.

Besides the most known way of generating lattice Gribov copies 
based on 
the change of the gauge fixing
process starting configurations
 by random gauge transformation, it is quite easy to 
produce Gribov copies on the lattice. 
As an example of that, we show in Fig.~\ref{F_over}
the lattice Gribov copies produced by the overrelaxation 
method~\cite{papar92} when varying the overrelaxation parameter 
$\omega$.
In Fig.~\ref{F_over}, different choices for the $\omega$ 
parameter, besides changing the   the gauge 
fixing algorithm rate of convergence,
lead to different Gribov copies even if starting from the same
initial configuration.
Of course there is no correlation between the convergence 
rate and the value of $F$ associated with the particular lattice
Gribov copy found.
\begin{figure}      
\hspace{2.0cm}  
\ifig{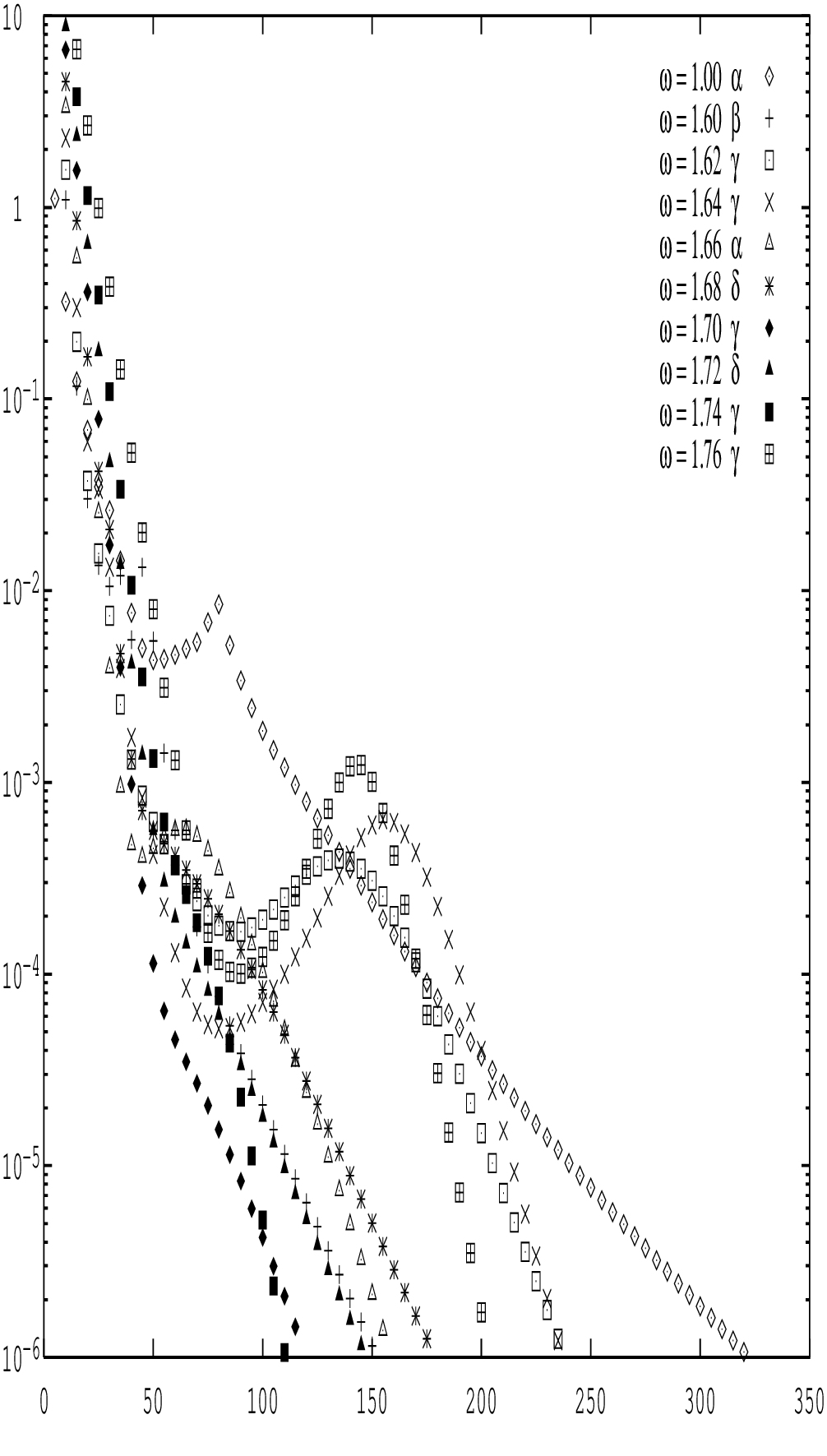}  
 \caption{$\theta$ as a function of the Landau gauge fixing 
sweeps for an $SU(3)$ lattice for different values of the 
overrelaxation  parameter $\omega$. The same letters $(\alpha, 
\beta, \gamma, \delta)$ indicates curves with the same final 
$F$ value. (from Ref.~\cite{papar92})} 
 \label{F_over}
\end{figure}
This can be qualitatively understood by recalling that non 
linear dynamical systems (and a gauge fixing algorithm is 
equivalent to such a system) often exhibit chaotic behaviour, 
in the sense that their evolution may depend dramatically 
both on the initial conditions (in this case the different 
random initial gauge rotations) and on the relevant parameter 
of the evolution equations (the overralaxation $\omega$ parameter).

To remove the Gribov ambiguity it has been proposed 
to fix first to a uniquely defined axial gauge \cite{manduogi90}.
But even if the starting gauge field is unique, one can still 
run into different local minima by using different gauge fixing 
algorithms. Moreover the axial gauge breaks rotational invariance, and 
some effect of this might be present in the local minimum that 
is favoured by this particular starting configuration.

\subsection{ Gribov Copies and Measurements}
\label{subsec:noise}
 
The numerical effects of lattice Gribov copies can be roughly divided 
into two categories: the distortion of a measurement and 
the lattice Gribov noise~\cite{petra}. 
The typical example of a distortion due to the existence of
Gribov copies is the measure of the photon propagator in 
compact $U(1)$ in the so called Coulomb phase. In this case 
the measure of the photon propagator as a function of the momentum, 
performed using the gauge fixing in the standard way, is 
affected by a not regular behaviour \cite{plewnia,defo91,mitri93}.
This problem has been associated with the distortional effects 
due to the Gribov copies. In fact, after having chosen the 
gauge fixed configurations nearest to the minimum of the gauge 
functional, the photon propagator became a smooth, regular
momentum function. More recent studies  \cite{mitrj} show
the details  of the Gribov copies dynamics within the Lorentz 
gauge and provide a practical procedure to eliminate their effects
in compact $U(1)$ physics.
 In fact, in the Coulomb phase, 
it is shown that, apart from double Dirac sheets~\cite{mitri0},
 all Gribov copies originate 
mainly from the zero-momemtum modes of the gauge field~\cite{mitri1}. The 
removal of these modes turns out to be necessary for reaching 
the absolute maximum of the gauge functional.

It is well known that Gribov copies have a relevant r\^ole in the 
lattice numerical studies on the confinement vortex picture.
A local iterative minimization  in  the direct maximal center 
gauge (discussed in  the Section~\ref{subs:MCG})
selects every 
possible minima and each of these Gribov copies has its 
own set of P-vortices, which may show  different 
properties: numerical demonstration of the Gribov copies effects
 has been shown in Ref.~\cite{koto2}.
Starting from configurations fixed in the Lorentz gauge and then 
going to the maximal center gauge fixing one obtains, on average, a maximun 
higher than starting from random gauge and then 
applying the same gauge fixing procedure.
The resulting picture in the two 
cases is dramatically different. In the latter case the 
very close numerical agreement of the center projected string 
tension $\sigma_{Z(2)}$ of \cite{deldebfa97,deldebfa98} 
with its currently accepted value 'unprojected'  $\sigma_{SU(2)}$
is reproduced. In the former case there is essentially complete 
loss of information about the string tension value. 

Moreover, careful studies of the gauge copies problem for the direct center 
projection in $SU(2)$ have been performed in Ref.~\cite{borko20},  
indicating that this gauge is not appropriate for the investigation 
of the center vortices and that the Gribov copies problem is more severe 
than it was thought before. In ~\cite{borko20},
following a procedure defined in Ref.~\cite{bashi96}, a gauge 
dependent quantity $X$ is computed on the gauge copy corresponding 
to the highest maximum of the functional $F$,  $F_{\it{max}}(N_{\it{cop}})$ 
after having generated $N_{\it{cop}}$ gauge equivalent copies, 
for a given starting configuration. 
Averaging over statistically independent gauge field configurations
and varying $N_{\it{cop}}$ the function $X(N_{\it{cop}})$ is 
obtained and extrapolated to $N_{\it{cop}} \to \infty$ limit. 
This procedue was built up in order to approach  the global 
maximum as close as possible. The effect of the Gribov copies 
is clearly shown drawing  the values of the $Z(2)$-projected 
Creutz ratios $\chi_{Z(2)}(I)$ as a function of $N_{\it{cop}}$ for 
$\beta = 2.5, L^4 = 16^4$ plotted in Fig.\ref{ncopies}.
$\chi_{Z(2)}(I)$ is defined through the projected Wilson loops 
$W_{Z(2)}(C) = exp\{i\pi{\cal L}(\sigma_P,C)\}$ where 
${\cal L}(\sigma_P,C)$ is the $4D$ linking number of the 
closed surface, $\sigma_P$, formed by P-vortex and closed 
loop $C$ 
\begin{equation}
\chi_{Z(2)}(I) = - log\frac{W_{Z(2)}(I,I)W_{Z(2)}(I+1,I+1)}
                     {W_{Z(2)}(I,I+1)W_{Z(2)}(I+1,I)}\;
\end{equation}
The Creutz ratios $\chi_{Z(2)}(I)$ are used to estimate 
the projected string tension $\sigma_{Z(2)}$.
\begin{figure}      
\ifig{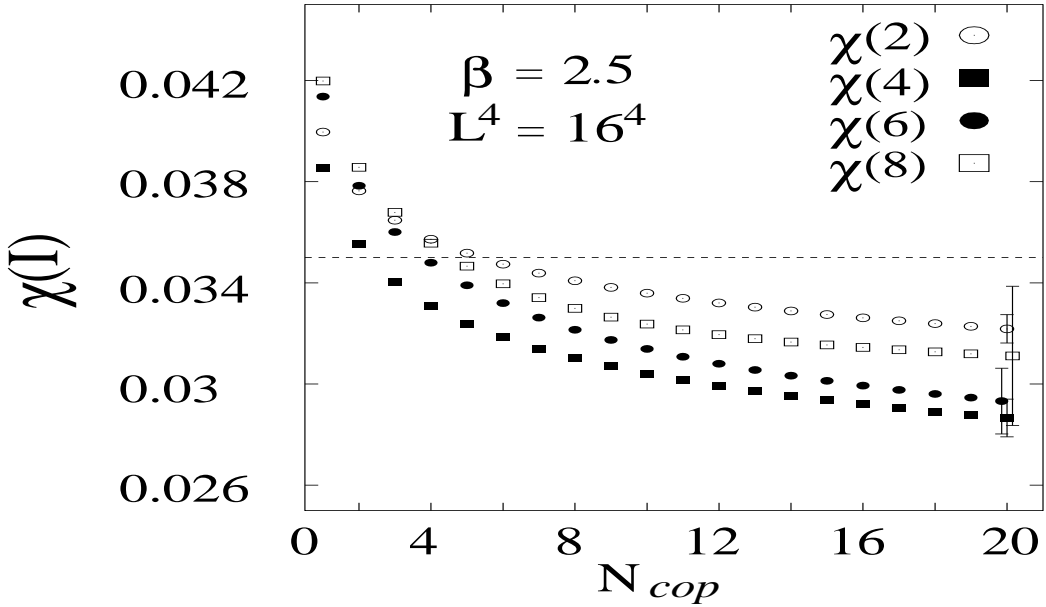}  
 \caption{The dependance of the Creutz ratios  $\chi_{Z(2)}(I)$ on the 
     number of gauge copies $N_{\it{cop}}$ for $\beta = 2.5$ 
    as obtained in Ref.~\cite{borko20}}  
 \label{ncopies}
\end{figure}
An update of the "drama of 
Gribov copies" on the center vortices studies can be found in 
Ref.~\cite{borko21}. In this paper  the disagreement 
between the projected string tension $\sigma_{Z(2)}$ in the direct 
maximal center gauge and the physical string tension $\sigma_{SU(2)}$
is demonstrated. 
The 
Laplacian center gauge~\cite{aledefo99}, 
appearing naturally as an extension of Laplacian 
Abelian gauge, is another attempt to get rid of Gribov ambiguity.

It is interesting  to note that in the 
measurements of the gluon propagator in $SU(3)$ there is no  
signal  about the Gribov ambiguity (for a 
recent review see Ref.~\cite{mandula99}).
The numerical simulations are performed in the Landau 
gauge and the various authors claim that the effects of 
Gribov copies do not affect the measurement~\cite{cucc}.
 Anyway, in the more regular case in which there is no distortion
due to Gribov copies, there  should  be  an increase of 
the numerical fluctuations due to the incomplete gauge 
fixing associated with the copies.

An attempt to study the properties of this noise has 
been done in Ref.~\cite{za0,za,coparpet95} taking as an example the 
measurement of the lattice axial current $Z_A$. This quantity
is particularly well suited to the Gribov 
fluctuations study. In fact $Z_A$ is a gauge independent quantity 
which can be obtained from chiral Ward identities in two 
distinct methods: a gauge independent one,
based on the matrix elements between hadronic states, and a gauge 
dependent one,  based on the matrix elements 
between quark states in the Landau gauge.
In  the intermediate steps of the numerical computation, 
the second procedure takes into account gauge dependent 
matrix elements potentially subjected to the Gribov noise.
Hence, there is an explicitly gauge invariant estimate 
of $Z_A$ which is free of Gribov noise and which can be 
directly compared to the gauge dependent, Gribov affected, 
estimate.
The results of the analysis~(\cite{za0,za}) can be summarized in the 
following way:
\begin{itemize}
\item  there is a clear evidence of residual gauge freedom 
 associated with lattice Gribov copies;
 \item the lattice Gribov noise is not separable from the 
statistical uncertainty of the Monte Carlo method.
\end{itemize}
The global effect is not dramatic because the $Z_A$ value 
obtained with the gauge dependent methods is close to the gauge 
independent evaluation  and the jacknife errors are comparable.

The influence of the Gribov copies has been studied also in 
the Coulomb gauge on the smeared correlation functions that 
are involved in the $B$ meson decay constant determination. 
The $B$ meson physics on the lattice \cite{papape92}, 
involving gauge dependent operators like ``smeared'' sources 
for quark correlation functions, is a suitable sector for 
an investigation of the Gribov ambiguity. In fact, if the 
gauge condition implemented numerically (e.g. the Coulomb gauge) 
does not correspond to a complete gauge fixing, in the sense 
that the gauge fixing algorithm may converge randomly to 
any configuration in a set of Gribov copies , then the 
value of the operators will depend on which copy gets 
selected by the algorithm. Then such residual gauge freedom 
acts as a source of statistical noise in the Monte Carlo 
average of those physical quantities that are extracted from 
gauges dependent quantities.

It must be noted that the Coulomb gauge condition (\ref{eq:effel})
is implemented indipendently on each timeslice, given a 
starting link configuration. This is because the links in 
the time direction which connect adjacent timeslices do not 
appear in the definitions of $F_U[G]$. In other words, each 
timeslice of a given configuration is endowed with its own 
scenario of Gribov copies, so that the pattern of their 
occurrence in the Coulom gauge is richer than in the Landau 
gauge, in which they are defined globally on the whole lattice.
One finds that the residual gauge freedom associated to Gribov 
ambiguity induces observable noise effects, though at the level 
of numerical accurary of considered simulation these effects are 
not relevant to the final determination of $\it f_B$. 
The results in Ref.~\cite{papape92}, obtained from $SU(3)$ lattice 
configurations generated on a $10^3 \times 20$ lattice indicate that 
such effects may become important on bigger lattices. In fact, 
increasing the lattice size would typically reduce the standard 
statistical noise, allowing, in principle, to perform measurements 
from large time separations, but then the gauge noise effect 
(which is not expected to disappear in the continuum 
limit) may become a major source of fluctuations and may provide 
a relevant contribution to the error bars on the physical 
quantities evaluated from smeared correlations.

Many numerical studies of Gribov ambiguities 
on abelian observables on the lattice  have 
been performed in the last years. In the studies of 
the dual superconductor hypothesis of confinement  one 
mainly uses the lattice version \cite{kronfeld87} of 't Hooft 
maximally Abelian gauge \cite{thooft81} sketched in 
Section \ref{subs:MAG}. The effect of Gribov copies is clearly seen
but the gauge fixing seems to be under 
control~\cite{bashi96,bapa95,harte97,bruckhei20}.
In particular, a careful study of the Gribov ambiguities in dual
superconductor scenario has been performed in Ref.~\cite{bashi96}.
 A new effective algorithm (simulated annealing~\cite{bapa95}
complemented with overrelaxation)  fixing maximal abelian gauge
to reduce gauge 
fixing ambiguities with respect to the standard overrelaxation 
algorithm and a numerical procedure to estimate the remaining 
gauge fixing uncertainties are used in this analysis.
It is important to emphasize that, using a new method to assess uncertainties
due to the incomplete gauge fixing, 
a procedure has been suggested to extrapolate values obtained on local 
maxima to the absolute maximum; the accuracy of all computations 
is limited by the statistical error on the biases.
The investigation of Ref.~\cite{bashi96} revealed that the effect of 
gauge copies cannot be neglected with respect to the statistical errors.
Nevertheless, the proposed algorithm does reduce the variance 
of observables versus the gauge copies in a sizeable way and
yields larger values of the functional to be maximized.
Then the results for the string tension show that its abelian part 
accounts for $92$ $\%$ of the confinement part in the static 
potential (on a $32^4$ lattice at $\beta = 2.5115$). 
 
By analogy with the Laplacian gauge fixing a Laplacian Abelian 
gauge~\cite{dersi97} (LAG), for the abelian projection, has been introduced.
Also this gauge shares the smoothness properties with the Landau gauge
but avoids lattice Gribov copies.  
This gauge fixing, which follows the construction of the 
Laplacian gauge, is based  on the lowest-lying eigenvector
of a covariant laplacian operator.  A numerical study on abelian 
and monopole dominance in MAG and in LAG has been performed in 
Ref. \cite{dersi99}. An investigation and a comparison  of the 
monopole structure of the $SU(2)$ vacuum as seen in different 
gauges are also performed in Ref.~\cite{ilge99}.

  \section{Gauge Dependent Smoothing Methods}  
\label{sec:SMOO}
It is known that for the matrix elements of local operators the
noise-to-signal  ratio diverges as the continuum limit 
is approached (as discussed for example in 
Ref.\cite{PariCarg83}).
In particular, to check the scaling of physical quantitites, 
their dependence over 
$\beta$ at the critical point of the theory has to be
computed . Then it is 
necessary to go to higher  $\beta$ values, 
despite the fact that the lattice spacing  shrinks 
when $g^2 \rightarrow 0$. Operators of a fixed lattice 
extent become smaller in physical size, and their
projection over the ground state goes to zero. A possible
trick in order to get a 
non-vanishing signal can be to resort to non local operators. 
Moreover, some kind of smoothing procedure 
is needed in order to study on the lattice the monopole-related 
observables and to determine the global topological charge in an 
unambiguous way\cite{ilge86}. 
The smoothing procedures kill part of the gauge field degrees of 
freedom in the process and these algorithms are not 
full gauge fixing methods. 

Several methods have been introduced 
to remove the unphysical short-distance fluctuations: {\em cooling} 
\cite{berg,perez9699}, {\em smearing} \cite{Falcio85,Ape87,Mari88}, 
{\em fuzzy loops} \cite{Te87,Degra87,Mite88} and other
renormalization group smoothing methods (see for example 
Ref. \cite{ilge99} and reference therein). 
Here we will give just a brief  description of smearing 
and  fuzzy loops procedures. For the smearing we
will discuss the gauge dependent procedure and for the  cooling methods
we send back to the quoted literature. 

The smearing procedure as originally proposed for $SU(3)$ 
in \cite{Falcio85} consists in the construction of 
correlation functions for operators which are a functional 
of the field smeared in space and not in time.  Then smeared operators, 
usually used for the QCD spectroscopy and phenomenology 
on the lattice depend on the gauge and then it is necessary 
to fix the gauge before their measure. 
For each link of a configuration  
the product of the other three links 
defining a plaquette
is considered , then  these products are summed over the four 
choices of plaquettes orthogonal to the time axis; 
the resulting matrix, projected on the gauge group, is
 the new link variable. A graphic description \cite{VANKRO} 
of the procedure is given in  Fig.\ref{F_SMEA1}. 
\begin{figure}      
\hspace{2.0cm}
\begin{center}   
 \includegraphics[height=20mm,width=70mm] {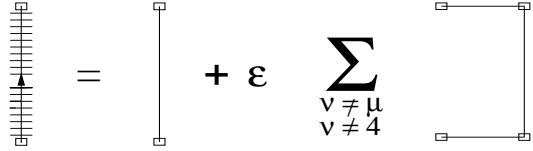} 
\end{center}
  \caption{Description of Ape's algorithm for smearing from
  Ref.~\protect\cite{VANKRO}.} 
  \protect\label{F_SMEA1}
\end{figure}
The value of the smearing coefficient $\epsilon$ is tuned in
order to  optimize 
the performance of the method \cite{Ape87}.
This procedure can be iterated to obtain a field more and 
more smeared in space at fixed time. 
Then one can have a set 
of operators, to implement a variational calculation, 
allowing to decide if a given euclidean time distance 
is asymptotic. Otherwise, as the amplitude of different 
exponential contributions depends on the operator, the ground 
state mass estimate can depend on the choice of the operator.
In the momentum space the effect of the smearing is equivalent 
to the application of  the factor $\exp({-k^2})$ (as it is 
shown in Ref. \cite{Ape87} for the example of a scalar field). 
This consideration makes more explicit the r\^ole of 
of short  wavelength fluctuations suppressor taken by the smearing.
Variations of the APE-style smearing have also been proposed in 
Refs. \cite{Guptapre,bonnetfitz20}. 

The procedure of smearing may be used to improve any local 
operator involving link variables. Smearing methods to improve 
lattice estimators have been already widely employed in the 
study of long distance correlations, such as large Wilson 
loops and hadron source operators. Just to give some  explicit 
example we  cite the improved lattice topological charge 
density operators, the hybrid (gluons as constituents) and 
glueball source operators. 
The first, constructed by a smearing-like 
procedure \cite{digiacomo1,digiacomo4,digiacomo5}, bears a 
better statistical behaviour as estimator of the topological 
density on the lattice both of $SU(2)$ Yang-Mills theory 
and of the full QCD. For the second, it is necessary to use ``smeared''
or ``fuzzed'' links for the gluon parts of the  operator to get 
good overlap with the ground state particles \cite{tou20}
.
To increase the overlap of hadron operators with their ground states
the creation of a smeared quark source has been proposed in
Ref.\cite{gusken1,gusken2}. This source, defined as
$S(\vec{x}) =  (D^2 + m_{sc}^2)_{3d}^{-1}\delta_{\vec{x},0}$
 where
the parameter $m_{sc}$ was tuned to give the quark source an
r.m.s. radius of about $3$ lattice spacing, has been used for the
the construction of smeared quark propagators.

A smearing procedure, described in Ref.~\cite{Blum:1997uf}, has 
been proposed to overcome the problem of Gribov copies on SU(2) 
lattices \cite{defor98}. Following this procedure, for each link
variable $U_{\mu}$ the sum $R_{\mu}$ of the 6 connecting staples 
is computed and $U_{\mu}$ is replaced with the combination:
\be
U_{\mu}^{s}=(U_{\mu}+wR_{\mu})/(1+6w)
\ee
where $w$ is an adjustable parameter and $U_{\mu}^{s}$ is reunitarized.
The procedure can be iterated many times; it turns out numerically 
that there is a critical value $w_c$ of $w$ below which the 
average plaquette goes to 1 (completely frozen lattice) for large
values of the number of the smearing steps.
The method is based on the following observation: when a
large number of smearing steps are performed, the so called 
``trivial orbit'' 
can be approached obtaining an unique
 Landau gauge configuration.  
Applying this gauge transformation to the original configuration, 
a unique starting point on the physical orbit is reached being the 
gauge path from this point to Landau point unambiguous.
The smearing is stopped when a sufficiently frozen lattice is obtained.
The method has been tested in $SU(2)$ theory
at $\beta=2.00,1.75,1.50$. 

A different smoothing method, the {\em fuzzy loops} procedure, is 
depicted~\cite{VANKRO} in  Fig.\ref{F_SMEA2}.
\begin{figure}        
\hspace{3.0cm}
\begin{center}  
 \includegraphics[height=30mm,width=90mm]{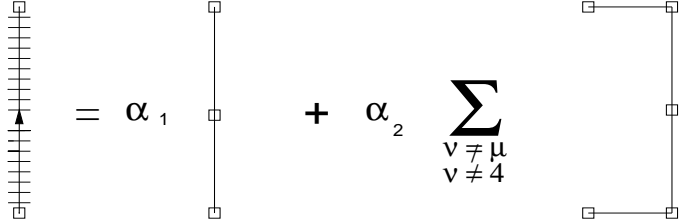}
\end{center}  
  \caption{Description of Teper's algorithm for fuzzy links 
from  Ref.~\protect\cite{VANKRO}.}
  \protect\label{F_SMEA2}
\end{figure}
In principle the $\alpha _i$ coefficients are tunable even 
if $\alpha _1=\alpha _2$ is chosen \cite{Mite88} 
to normalize the fuzzy link back into $SU(3)$.
A consequence of fuzzy technique is that {\em links paths} 
grow exponentially during the iterations and there are $2^3$ 
times fewer fuzzy links than original links. Then, after 
the iteration, a simple loop of fuzzy links is a complicated 
linear combination of loops of original links. Finally 
elementary loops of fuzzy links are quite non local when 
expressed in terms of the original links. 
We note that the fuzzy procedure is inspired by the Monte 
Carlo renormalization  group \cite{Wil80,Sheto80,Swe81} 
methods involving factor-of-two blocking. 
Another renormalization group based smoothing method 
\cite{degrahase96} is employed \cite{ilge99} on the lattice 
to investigate the monopole structure of the $SU(2)$ vacuum. 
It is suitable to eliminate UV lattice artifacts from the 
monopole-related observables without destroing the confining 
structure as other procedures finally do.

Recently a new {\em cooling} approach \cite{perez99} has been 
used as a gauge invariant low pass filter to extract physical 
information from noisy Monte Carlo configurations.
In Ref.~\cite{bonnetfitz20} Wilson action and topological 
charge are used to determine the relatives rates of standard 
cooling and smearing algorithms in pure $SU(3)$ color gauge 
theory. 

\section{Final Remarks and Acknowledgements}
\label{sec:CONCLUSIONS}

The purpose of this paper is to collect many results and 
considerations 
about numerical gauge fixing on the lattice 
which are usually scattered in the literature.
We selected the topics on the basis of our interests and competence
and we apologize for the many subjects either we have
just sketched or we have not discussed at all.

We  have tried to give, in a clear way, an overview of
the problems due to the definition
of the gluon field and to the incomplete Landau gauge fixing as it is
performed on the lattice today. 

We thank for many discussions: 
Giuseppe Marchesini and Francesco Di Renzo about the gauge fixing in the Langevin scheme, 
Enzo Marinari
about the gauge dependent smoothing methods and Valya~Mitrjushkin
about the Gribov copies in compact $U(1)$ theory.
We warmly thank Massimo~Testa for many illuminating discussions and suggestions 
and for the careful reading of the manuscript.
S.~P. thanks the CERN theory division for the hospitality during the 
completion of this paper.
L.~G. has been supported in part under DOE grant DE-FG02-91ER40676.

\nonumsection{References}

\end{document}